\documentclass[prd,aps,floats,twocolumn]{revtex4}
\usepackage[usenames, dvipsnames]{color}
 \usepackage{graphicx, array}
\usepackage{graphics,epsfig}
 \usepackage{amssymb}
 \usepackage{amsmath}
 \usepackage{ulem}
 \usepackage{multirow}
  \usepackage{subfigure}
  \usepackage{ulem}

 \def\be{\begin{equation}}
\def\ee{\end{equation}}
 \def\bi{\begin{itemize}}
 \def\ei{\end{itemize}}
  \def\ben{\begin{enumerate}}
\def\een{\end{enumerate}}
  \def\bt{\begin{tabular}}
\def\et{\end{tabular}}
\def\bc{\begin{center}}
\def\ec{\end{center}}

\def\bea{\begin{eqnarray}}
\def\eea{\end{eqnarray}}

\def\ba{\begin{eqnarray}}
\def\ea{\end{eqnarray}}

\def\tr{\tilde{r}}
\def\bxi{\bar{\xi}}
\def\Vij{V}

\begin{document}

\input{epsf}

\title{Constraints on gravity and dark energy from the pairwise kinematic Sunyaev-ZelÕdovich effect}
\author {Eva-Maria Mueller$^{1,2}$, Francesco de Bernardis$^{1}$, Rachel Bean$^{2}$, Michael Niemack$^{1}$.}
\affiliation{ $^1$ Department of Physics, Cornell University, Ithaca, NY 14853, USA,
\\
$^2$ Department of Astronomy, Cornell University, Ithaca, NY 14853, USA.}
\label{firstpage}

\begin{abstract}

We calculate the constraints on dark energy and cosmic modifications to gravity achievable with upcoming cosmic microwave background (CMB) surveys sensitive to the Sunyaev-Zeldovich (SZ) effects.
The 
 analysis focuses on using the mean pairwise velocity of clusters as observed through the kinematic SZ effect (kSZ), an approach based on the same methods used for the first detection of the kSZ effect, 
 and includes a detailed derivation and discussion of this statistic's covariance
under a variety of different survey assumptions.

The potential of current, Stage II, and upcoming, Stage III and Stage IV, CMB observations are considered, in combination with contemporaneous spectroscopic and photometric galaxy observations.
A detailed assessment is made of the sensitivity to the assumed statistical and systematic uncertainties in the optical depth determination, the magnitude and uncertainty in the minimum detectable mass, and the importance of pairwise velocity correlations at small separations, where non-linear effects can start to arise.  

In combination with  Stage III constraints on the expansion history, such as those projected by the Dark Energy Task Force,  we forecast 5\% and 2\% for fractional errors on the growth factor, $\gamma$, for Stage III and Stage IV surveys respectively, and 2\% constraints on the growth rate, $f_g$, for a Stage IV survey for $0.2<z<0.6$. The results suggest that  kSZ measurements of cluster peculiar velocities, obtained from cross-correlation with upcoming spectroscopic galaxy surveys, could provide robust tests of dark energy and theories of gravity on cosmic scales.

\end{abstract}

\maketitle

%%%%%%%%%%%%%%%%%%%%%%%%%%%%%%%%%%%%%%%%%%%%%%
\section{Introduction}
\label{sec:intro}
%%%%%%%%%%%%%%%%%%%%%%%%%%%%%%%%%%%%%%%%%%%%%%

The accelerating expansion of the universe continues to be one of the most puzzling problems in cosmology. The background evolution of the universe is constrained by measurements of the cosmic microwave background (CMB) (e.g. \cite{Ade:2013zuv, Hinshaw:2012fq, Calabrese:2013jyk,Story:2012wx}), baryon acoustic oscillations (BAO) in the galaxy two point correlation function (e.g. \cite{Eisenstein:2005su,Percival:2007yw,Percival:2009xn,Kazin:2014qga}),  as well as  type 1a supernovae (SN) (e.g. \cite{Riess:1998cb,Perlmutter:1998np}).

There is significant interest in differentiating between alternative explanations of cosmic acceleration by extending beyond the expansion history to dark energy's impact on the growth of structure (see \cite{Uzan:2010ri,Jain:2013wgs,Joyce:2014kja} for reviews).
This approach is key if a modification of gravity on astrophysical scales is responsible for cosmic acceleration. Large scale structure observations provide two complementary probes of the properties of gravity: the
  bending of light due to a gravitational potential
and the effect of gravity on the motions of non-relativistic objects. The latter manifests as the peculiar velocities of galaxies imprinted in redshift space distortions (RSD) in the galaxy correlation function \cite{Huterer:2013xky} as well as cluster motions as observed through the kinematic Sunyaev-Zel'dovich (kSZ) 
effect \cite{Sunyaev:1980nv}. Upcoming surveys such as the Dark Energy Survey
 (DES)  \citep{Abbott:2005b}, HyperSuprimeCam (HSC) \footnote{http://sumire.ipmu.jp/en/}, the Large Synopic Survey Telescope (LSST) \cite{Abell:2009aa} and the Euclid \cite{Laureijs:2011gra} and WFIRST \cite{Spergel:2013tha}
space telescopes, will provide gravitational lensing surveys out to redshift 2, and beyond. Concurrently spectroscopic surveys such as eBOSS \citep{Comparat:2012hz}, DESI \cite{Levi:2013gra} and spectroscopy from Euclid and WFIRST,  will provide both BAO and redshift space clustering  measurements over overlapping epochs and survey areas. 
Each of those probes, though having the potential to constrain gravity, are affected by systematic effects. Cosmological measurements using weak gravitational lensing (WL) will require precise photometric redshift and point spread function calibrations along with characterization of  intrinsic alignment contamination of shear correlations, e.g. \cite{Abate:2012za}, that can bias and dilute dark energy constraints \cite{Joachimi:2009ez,Laszlo:2011sv,Kirk:2011sw}. Accurate modeling of redshift space clustering into the non-linear regime requires precise descriptions of the galaxy clustering correlations beyond the Kaiser formula \cite{Kwan:2011hr}.  Clusters are high density environments that are highly affected by the underlying theory of gravity. The peculiar velocities of clusters provide an alternative, complementary measurement of the cosmological gravitational potential field that has different systematic uncertainties. Considering these as part of a multiple tracer approach will provide the clearest picture of gravity's properties.

Cluster motions leave a secondary imprint in the CMB known as the 
kSZ effect \cite{Sunyaev:1980nv}, the process of CMB photons passing through a cluster and being Doppler shifted due to the cluster's peculiar velocity relative to the CMB rest frame. This provides a potentially powerful complementary measurement of gravity's influence on cosmic structure to the peculiar motions of individual galaxies \cite{Zhang:2004yt,Diaferio:2004gk,HernandezMonteagudo:2005ys,DeDeo:2005yr,Fosalba:2007bx,Bhattacharya:2006ke,Kosowsky:2009nc,2011ApJ...736..116M}.
Despite its potential,  the kSZ has been hard to measure; the signal is small when compared to the thermal SZ effect and emission from dusty galaxies, and doesn't have a distinct frequency dependence. Observational efforts to constrain the cluster peculiar velocities have come from multi-band photometry in combination with X-ray spectra \cite{1997ApJ...481...35H,2003ApJ...592..674B,2004PASJ...56...17K,2012MNRAS.421..224M,2012ApJ...761...47M} and spectroscopy around the thermal SZ null frequency \cite{2012ApJ...749..114Z}. Recent work extracted  the kSZ signature  from individual clusters by combining sub-mm, X-ray  and sub-arcminute resolution CMB data to respectively remove dusty galaxy emission, estimate electron density and fit thermal and kinematic SZ templates \cite{Sayers:2013ona}. Data from the {\it WMAP} and {\it Planck} satellites have been used to place upper limits on the bulk flows and statistical variation in cluster peculiar velocities \cite{Osborne:2010mf,Ade:2013opi}, while  South Polar Telescope (SPT) data \cite{2012ApJ...755...70R,2012ApJ...756...65Z} and Atacama Cosmology Telescope (ACT) data \cite{Sievers:2013wk} have been used to place limits on the kSZ signal from the epoch of reionization.

Multi-band methods 
do not yet provide a practical approach to extract the kSZ signal from thousands of clusters as desired for large scale cosmological correlations.  Cross-correlating arcminute resolution CMB maps with cluster positions and redshifts determined by a spectroscopic large scale structure survey 
can enable extraction of the pairwise kSZ signal \cite{Bhattacharya:2006ke,Kosowsky:2009nc, Li:2014mja}. 
Indeed, the first detection of the kSZ effect in the CMB spectrum was made by combining CMB measurements from the ACT \cite{fowler/etal:2007} with the SDSS BOSS spectroscopic survey \cite{Comparat:2012hz} to measure the mean pairwise momentum of clusters, using luminous red galaxies as a tracer for clusters  \cite{Hand:2012ui}. The pairwise approach for extracting the kSZ signal 
measures the difference in peculiar velocities of nearby clusters as a function of the comoving distance between the clusters. This approach minimizes contributions from the CMB, thermal SZ, and foregrounds, which can be treated as approximately constant on these scales, and by averaging over many clusters pairs any effects independent of the separation will cancel.
CMB surveys such as ACTPol \cite{Niemack:2010wz}, SPTPol \cite{Austermann:2012ga}, Advanced ACTPol \cite{Calabrese:2014}, SPT-3G \cite{Benson:2014}, and a next-generation, so-called Stage IV CMB survey \cite{CMBS4neu:2013} in combination with overlapping galaxy surveys, such as those described above,
can improve upon this detection and enable the use of mean pairwise velocities as a cosmological probe.

In this paper, we study the constraints on dark energy and cosmic modifications to gravity  expected from analyzing the mean pairwise velocity of clusters observed through the kSZ effect by upcoming 
CMB observations in combination with spectroscopic large scale structure redshift surveys.
In section \ref{sec:Formalism} the analytical formalism used to construct statistics and associated covariances for cluster velocity correlations is summarized. The analysis approach and findings are presented in section \ref{sec:analysis}, with conclusions and implications for future work discussed in section \ref{sec:conclusions}. A detailed derivation of key results  in \ref{sec:Formalism}  is presented in Appendix \ref{app:cov}.

%%%%%%%%%%%%%%%%%%%%%%%%%%%%%%%%%%%
%\section{Modeling observations}

%%%%%%%%%%%%%%%%%%%%%%%%%%%%%%%%%%%%%%%%%%%%%%
\section{Formalism}
\label{sec:Formalism}
%%%%%%%%%%%%%%%%%%%%%%%%%%%%%%%%%%%
We consider the mean pairwise velocity of clusters derived from the kSZ effect as a probe for dark energy models and modifications to general relativity. Section \ref{sec:MG_LSS} outlines how the growth of structure can be used to constrain modified gravity, \ref{sec:Vel} summarizes the halo model approach to analytically calculate the mean pairwise velocity of clusters, and \ref{sec:Cov} presents the formalism to estimate the covariance matrix of the mean pairwise velocity. In sections \ref{sec:cosmomodel} and \ref{sec:survey_spec} we discuss the fiducial cosmological model and survey assumptions. 

%%%%%%%%%%%%%%%%%%%%%%%%%%%%%%%%%%%%%%%%%%%%%
\subsection{Cosmic structure and modified gravity}
\label{sec:MG_LSS}
%%%%%%%%%%%%%%%%%%%%%%%%%%%%%%%%%%%%%%%%%%%%%
Even though on large scales the universe appears homogenous and isotropic, initial local matter overdensities form galaxies and galaxy clusters and evolve into the large scale structure of the universe.
The growth of these structures depends on the underlying physical theory and can therefore be used to constrain cosmological models.

According to linear theory, the matter over-density, $\delta_m$, is related to the velocity of dark matter particles, $\dot{\delta}_m \propto v_m$, 
which connects the time evolution of the perturbations to the
dark matter velocity.  Any tracer of the underlying dark matter velocity distribution can be used to constrain cosmology and in particular modified gravity models.
In a variety of modified gravity scenarios the evolution of the density perturbations can be quite different from standard gravity even though the background expansion of the universe is undistinguishable from a $\Lambda CDM$ universe (for example \cite{Laszlo:2007td,Amendola:2007rr,Linder:2007hg}).
The linear perturbation equations have a solution of the form $\delta_m(\vec{x},t)=D_a(t)\delta(x)$, factorizing the spatial and temporal dependency, with $D_a$ being the growth factor.
We can define the growth rate at a given scale factor, $a$, as
\bea
f_g(a) \equiv \frac{d\ln D_a}{d\ln a}
\eea
to parametrize the growth of structure. The growth rate is well approximated by $f_g(a)\approx \Omega_m(a)^\gamma$ with the growth index $\gamma \simeq0.55$ for standard gravity \cite{Wang:1998gt, Linder:2005in}.
Pairwise velocity statistics can be used to constrain the cosmological model of the universe and the underlying theory of gravity \cite{Juszkiewicz:1998xf,Ferreira:1998id}.

%%%%%%%%%%%%%%%%%%%%%%%%%%%%%%%%%%%%%%%%%%%%%
\subsection{Motion of clusters as a probe of cosmology}
%%%%%%%%%%%%%%%%%%%%%%%%%%%%%%%%%%%%%%%%%%%%%
%%%%%%%%%%%%%%%%%%%%%%%%%%%%%%%%%%%%%%%%%%%%%
\label{sec:Vel}
%%%%%%%%%%%%%%%%%%%%%%%%%%%%%%%%%%%%%%%%%%%%%
We analytically model the expected large scale motion of clusters under 
cosmological gravitational interactions by considering the properties of dark matter particles, in linear theory, and then using a halo model to  infer the velocity statistics of gravitationally bound halos, which we use as proxies for galaxy clusters.

%=================================================================
\begin{figure*}[!thb]
\bc
{\includegraphics[width=0.49\textwidth]{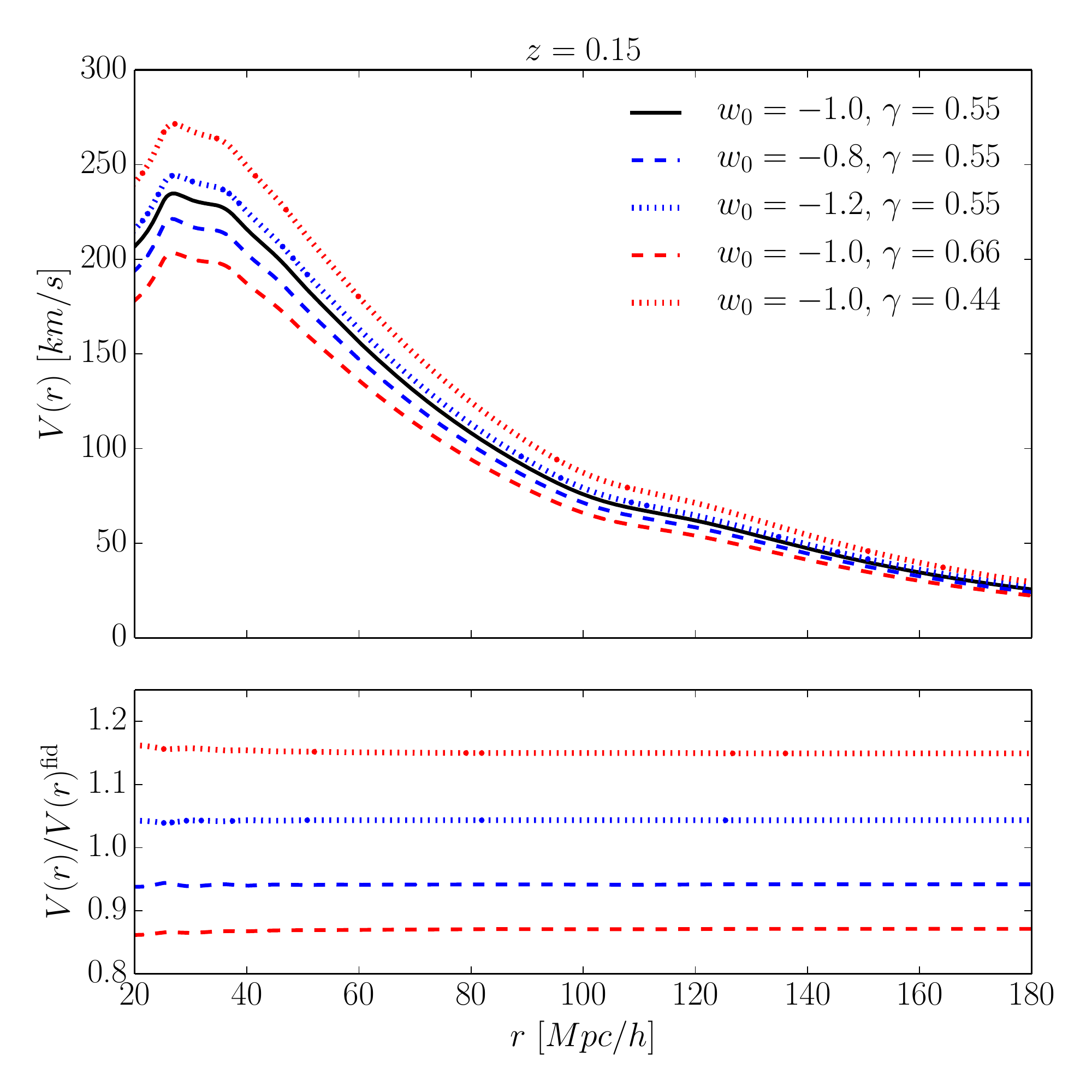}
\includegraphics[width=0.49\textwidth]{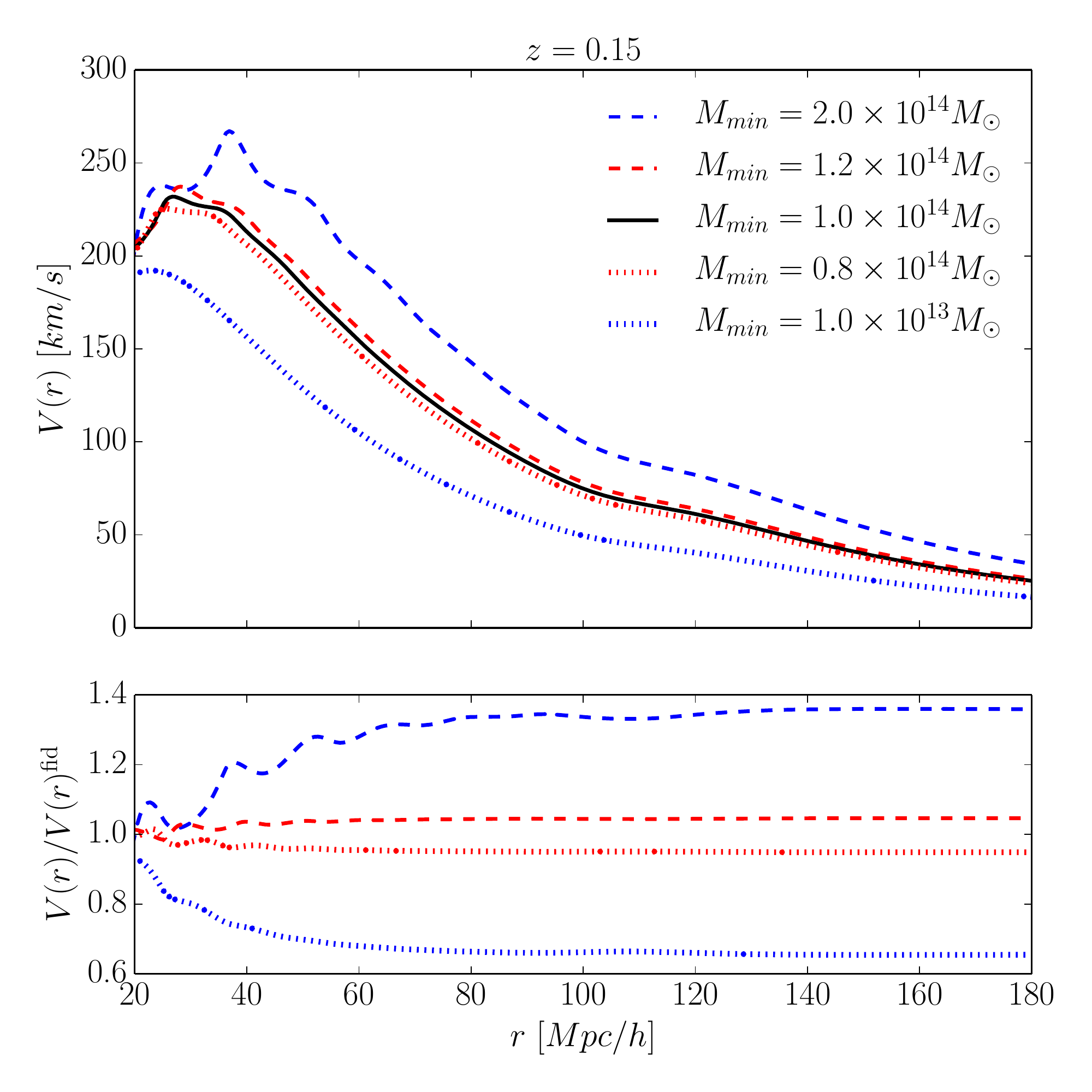}
}
\caption{[Left panel] The mean pairwise cluster velocity, $\Vij$, for different values of the dark energy equation of state parameter, $w_0$, and the modified gravity parameter, $\gamma$, at $z=0.15$ and assuming a minimum cluster mass of $M_{\mathrm{min}}=1\times 10^{14} M_\odot$. 
A more negative $w_0$ leads to an increase in $\Vij$. 
A decreased value of $\gamma$ increases the growth rate and therefore increases $\Vij$ whereas a higher value of $\gamma$ has the opposite effect. The same fractional change in $\gamma$ has a greater effect on the amplitude of $\Vij$  than changing $w_0$. [Right panel] The mean pairwise cluster velocity, $\Vij$, for different minimum mass cut-offs at redshift of $z=0.15$. Note that changing $M_{\mathrm{min}}$ changes the shape as well as the amplitude of $\Vij$. Higher $M_{\mathrm{min}}$ leads to an increase in mean pairwise velocity since the more massive clusters tend to have higher streaming velocities. [Lower panels] Ratio of the mean pairwise velocity, $\Vij$, for the different scenarios to that for the fiducial model, $\Vij^{fid}$. }    
\label{fig:Vij}
\ec
\end{figure*}
%=================================================================
Following the formalism outlined in \cite{ Sheth:2000ff}, we 
assume linear theory to describe the mean pairwise streaming velocity, $v$, between two dark matter particles, at positions $\bf{r}_i$ and $\bf{r}_j$, in terms of their comoving separation $r=|\bf{r_i}-\bf{r_j}|$,
\bea
v(r)&=&-\frac{2}{3} f_g(a)H(a)ar\frac{\bar{\xi}(r,a)}{1+\xi(r,a)}\label{eq:vij}
\eea
where $\xi$ is the dark matter 2-point correlation function and $\bar{\xi}$ the volume averaged correlation function, respectively defined as,
\bea
\xi(r,a)&=&\frac{1}{2 \pi^2}\int dk k^2 j_0(kr) P(k,a),
\\
\bar{\xi}(r,a)&=&\frac{3}{ r^3} \int_0^{r} dr' r'^2 \xi(r,a),
\eea
with $P(k,a)$ being the dark matter power spectrum and $j_0(x)=\sin(x)/x$ is the zeroth order spherical Bessel function.

The properties of dark matter halos of mass $M$,  relative to the dark matter distribution, can be modeled using a halo bias
\bea
b(M,z)&=&1+\frac{\delta_{\mathrm{crit}}^2-\sigma_0^2(M,a=1)}{\sigma_0^2(M,a=1)\delta_{\mathrm{crit}} D_a},
\eea
where $M(r) = 4\pi R^3\bar\rho/3$, $\bar\rho$ is the average cosmological matter density, the critical overdensity is taken to have the standard $\Lambda$CDM value of $\delta_{\mathrm{crit}} \approx 1.686$,
and  the zeroth order moment of the  mass distribution squared is
\bea
\sigma_0^2(m,a)&=&\frac{1}{2 \pi^2} \int_0^{\infty} dk k^2 P(k,a)W^2(kR(M)).
\eea

Surveys will generally include cluster halos over a range of masses above some limiting mass threshold, $M_{\mathrm{min}}$. To analyze the mass statistics we consider a {\it mass averaged} cluster pairwise velocity statistic, $\Vij$, for pairs of clusters separated by a comoving distance $r$
\bea
\Vij(r,a)&=&-\frac{2}{3} H(a) a f_g(a)\frac{r \bar{\xi}_{h}(r,a)}{1+\xi_{h}(r,a)}, \label{eq:massavvij} 
\eea
which has an analogous expression to that in (\ref{eq:vij})  
 \cite{Sheth:2000ff, Bhattacharya:2007sk}, with
\bea
\xi_{h}(r,a)&=&\frac{1}{2 \pi^2}\int dk k^2 j_0(kr) P_{\mathrm{lin}}(k,a) b_{h}^{(2)}(k), \hspace{0.5cm}
\\
\bar{\xi}_{h}(r,a)&=&\frac{3}{ r^3} \int_0^{r} dr' r'^2 \xi(r,a) b_{h}^{(1)}(k).
\eea
 The mass-averaged halo bias moments,  $b^{(q)}_{h}$, are 
given by
\bea
b^{(q)}_{h}&=&\frac{\int_{M_{\mathrm{min}}}^{M_{\mathrm{max}}} dM \ M \ n(M)b^{q}(M)W^2[kR(M)]}{\int_{M_{\mathrm{min}}}^{M_{\mathrm{max}}} dM\ M \ n(M)W^2[kR(M)]} \hspace{0.5cm} \label{eq:bq}
\eea
where $n(M)$ is the number density of halos of mass $M$, given by the Jenkins mass function, and the top-hat window function $W(x)=3(\sin x-x \cos x)/x^3$.

In Figure \ref{fig:Vij} we show the mean pairwise velocity, $\Vij$ as a function of cluster separation r for a number of  cosmological models at $z=0.15$ for a survey with limiting mass $M_{\mathrm{min}}=10^{14}M_\odot$, assuming all other survey specifications are fixed (left panel) and for various assumptions on the limiting mass (right panel). The figure suggests that, as with other linear growth rate related statistics,  the equation of state, $w_0$, and growth exponent, $\gamma$, have degenerate effects on the  pairwise velocity amplitude, through their effects on the growth factor, and do not alter the shape of the function. However, as indicated in section \ref{sec:measure_error}, the redshift dependence of these parameters helps to break the degeneracy. To be more specific, the amplitude of $V$ as a function of $z$ is different for variations in $\gamma$ compared to $w_0$. Increasing the minimum cluster mass shifts the peak of the pairwise velocity function to larger scales (on scales below 60 Mpc) and boosts the overall amplitude on scales larger than this, because the larger clusters have a larger streaming velocity.

%%%%%%%%%%%%%%%%%%%%%%%%%%%%%%%%%%%%%%%%%%%%%
\subsection{Covariance matrix}
\label{sec:Cov}
%%%%%%%%%%%%%%%%%%%%%%%%%%%%%%%%%%%%%%%%%%%%%
Measurements of cluster velocities are subject to a number of statistical and systematic uncertainties. First, discreteness effects need to be taken into consideration;
 a smooth continuous field is typically assumed to underly a discrete distribution of local objects, which leads to shot noise. For a large sample size the shot noise should be approximately Gaussian resulting in an error proportional to $1/N$ \cite{Eisenstein:1999jg}, where 
$N$ is the number of objects in the sample.
If the number of objects (e.g. clusters) in the sample is not sufficiently large, the Gaussian limit breaks down, and an additional non-Gaussian contribution to the shot noise can become relevant \cite{Cohn:2005ex}. 

Second, as in any cosmological survey, the measurement will be subject to cosmic variance due to the finite size of the sample. Third, in addition to the statistical errors we include a velocity measurement error 
\cite{Bhattacharya:2007sk} to account for the accuracy of the measurements and the uncertainty in the optical depth of the clusters. The total covariance for the mean pairwise velocity is therefore a combination of cosmic variance, shot noise, and the velocity measurement error:
\bea
C_{\Vij}^{\mathrm{total}}(r,r')&=&C_{\Vij}^{\mathrm{cosmic}}(r,r') + C_{\Vij}^{\mathrm{shot }}(r,r')  \nonumber 
\\
&&+C_{\Vij}^{\mathrm{measurement}}(r,r'). \hspace{0.6cm}\label{eq:cov}
\eea
A detailed derivation of the covariance terms can be found in Appendix \ref{app:cov}.  We summarize the results here.

Defining an estimator for the mean pairwise velocity, $\hat{\Vij}$, enables the covariance matrix to be calculated using
\bea
C_{\Vij}(r,r')=\langle \hat{\Vij}(r)\hat{\Vij}(r')\rangle-\langle \hat{\Vij}(r) \rangle \langle \hat{\Vij}(r') \rangle\label{eq:covdef}
\eea
where $\langle...\rangle$ is the volume average.
For analyzing a survey we include binning, as observations will be combined not at just one radius $r$ but in bins of width $\Delta r$,
\bea
\hat{\Vij}(r)\rightarrow V_{\Delta}(r)= \frac{1}{V_{\mathrm{bin}}} \int_{r-\Delta r/2}^{r+\Delta r/2} \tilde{r}^2d\tilde{r}  \int d\Omega  \hat{\Vij}(\tilde{r}),
\eea
assuming spherical symmetry and where a $\Delta$ subscript indicates binned quantities over bins of size $\Delta r$.

The covariance between the mean pairwise velocities of two cluster pairs, with the two pairs separated by $r$ and $r'$ and using bin width $\Delta r$,  can be expressed as
\bea
&&C_{\Vij}(r,r')=
 \frac{4}{\pi^2 V_s(a)} \left(\frac{H(a)a}{1+\xi_{h}(r,a)}\right)^2 f_g(a)^2 \hspace{0.5cm} \nonumber
 \\
 &&\times\left[\int dk \left( P_{\mathrm{lin}}(k,a) b^{(1)}_{h}(k)+\frac{1}{n(a)} \right)^2 W_{\Delta}(kr) W_{\Delta}(k r')  \right.\nonumber
 \\
 && +\left.   \hspace{0.35cm}  \frac{\Delta r}{V_{\Delta}(r')} \int dk k \frac{P_{\mathrm{lin}}(k,a) b^{(1)}_{h}(k)}{n(a)^2} W_{\Delta}(kr) \right],\hspace{0.5cm}   \label{eq:shot}
\eea
where $V_s(a)$ is the survey volume, and
\bea
W_{\Delta}(kr)&=&3\frac{R_{\mathrm{min}}^3 \tilde{W}(k R_{\mathrm{min}}) - R_{\mathrm{max}}^3 \tilde{W}(k R_{\mathrm{max}})}{R_{\mathrm{max}}^3-R_{\mathrm{min}}^3} \hspace{0.5cm}
 \\
  \tilde{W}(x)&=&\frac{2 \cos(x) + x\sin(x)}{x^3}.
\eea

\begin{figure*}[!tb]
\bc
{\includegraphics[width=0.48\textwidth]{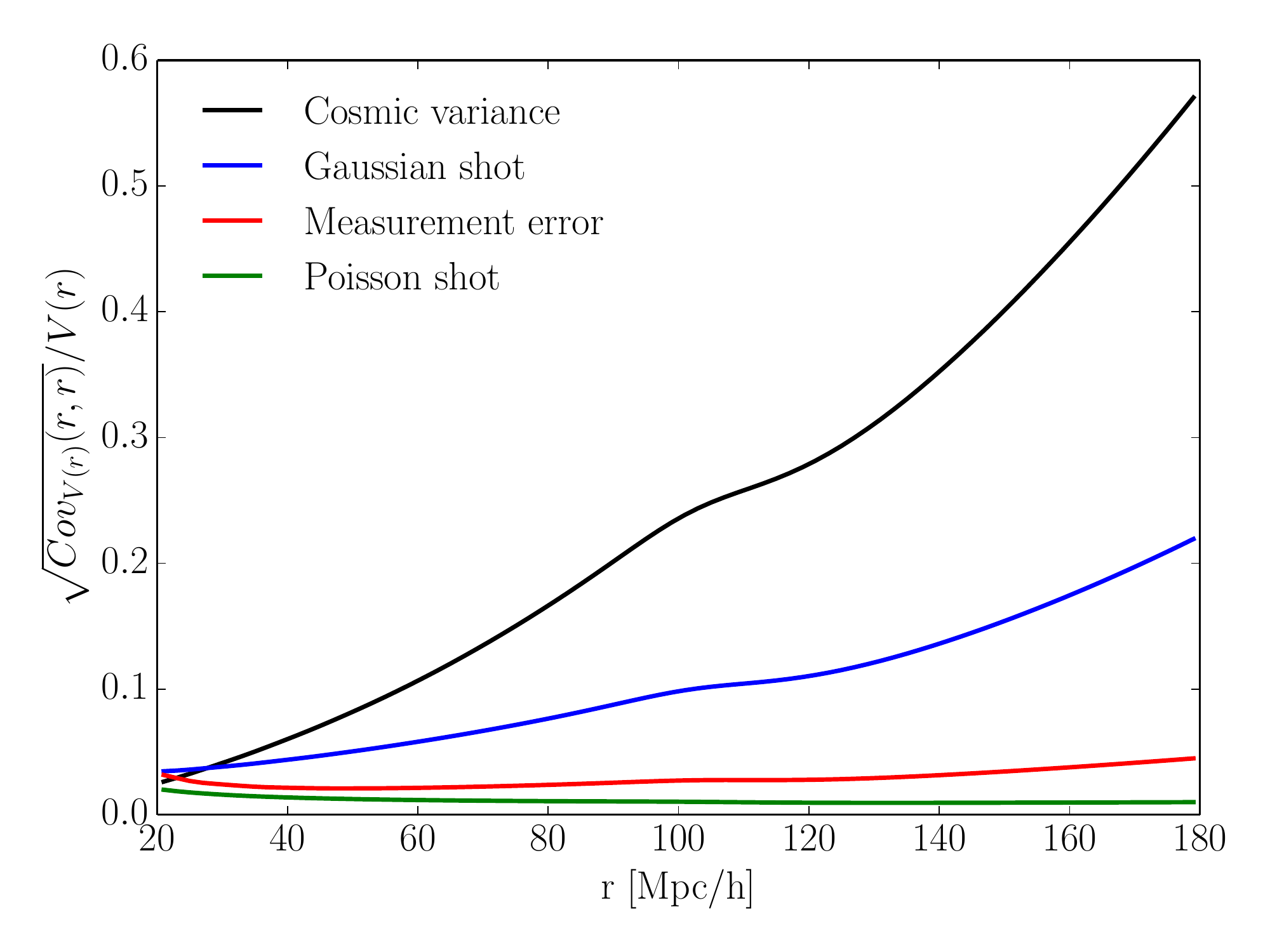}
\includegraphics[width=0.48\textwidth]{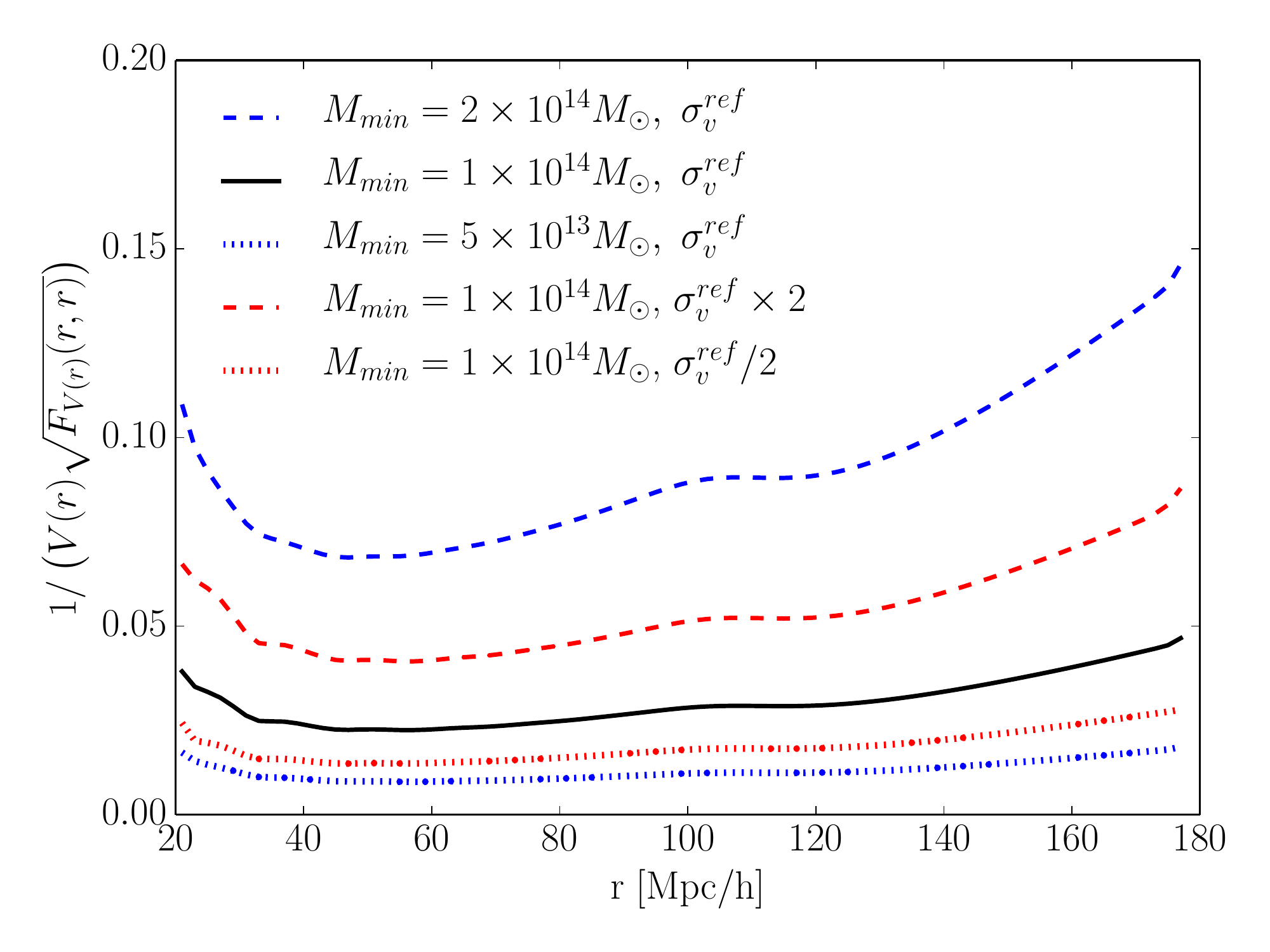}
}
\caption{[Left panel] The relative error on the mean pairwise velocity of clusters at redshift $0.1<z<0.2$ for a separation bin size $\Delta r=2 \ \mathrm{Mpc/h}$ assuming a Stage III like survey (see Table \ref{tab:kSZ_surveys}).  The Poisson shot noise is sub-dominant compared to the other terms, the Gaussian shot noise term however cannot be neglected.  [Right panel] One over the diagonal terms of the total Fisher matrix relative to the mean pairwise velocity for varying the minimum mass and the measurement error. The effect of the minimum mass on the fisher matrix is more prevailing than the dependency on the measurement error.}
\label{fig:CovBin}
\ec
\end{figure*}

The first term in (\ref{eq:shot}) is the Gaussian contribution to the covariance, which includes both cosmic variance ($\propto P$) and shot noise ($\propto 1/n$).  The second term is an additional contribution that is often neglected, which arises if the Gaussian limit breaks down; we refer to this term as `Poisson' shot noise as in \cite{Cohn:2005ex}. While we find it is subdominant in comparison to the Gaussian terms for a mass cut-off $M\leq 1\times 10^{14} M_{\odot}$ (see Figure \ref{fig:CovBin}), it can be important for surveys with smaller cluster number densities. The purely Gaussian shot noise contribution on the other hand is not insignificant and should be included.

%=================================================================
\begin{figure}[!tb]
\bc
{\includegraphics[width=0.5\textwidth]{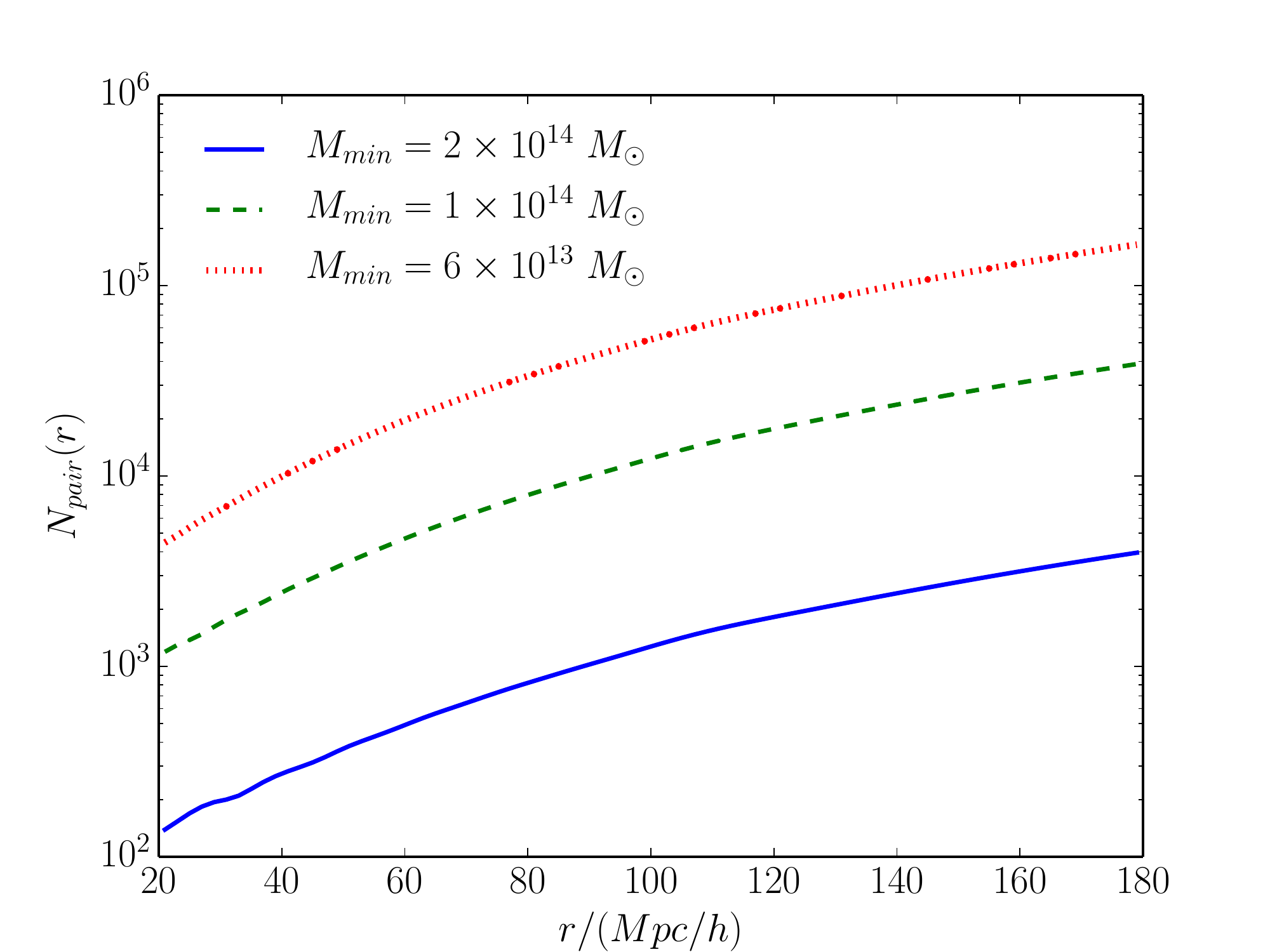}
}
\caption{Number of cluster pairs, $N_{\mathrm{pair}}(r)$, versus separation in bins of $\Delta r = 2$ Mpc/h for different mass cutoffs at redshift $0.1<z<0.2$.  This assumes a Jenkins mass function and a 6000 square degree survey.  
}
\label{fig:npair}
\ec
\end{figure}
%=================================================================

%=================================================================
%=================================================================

 %=================================================================
\begin{figure*}[!tb]
\bc
\includegraphics[width=.9\textwidth]{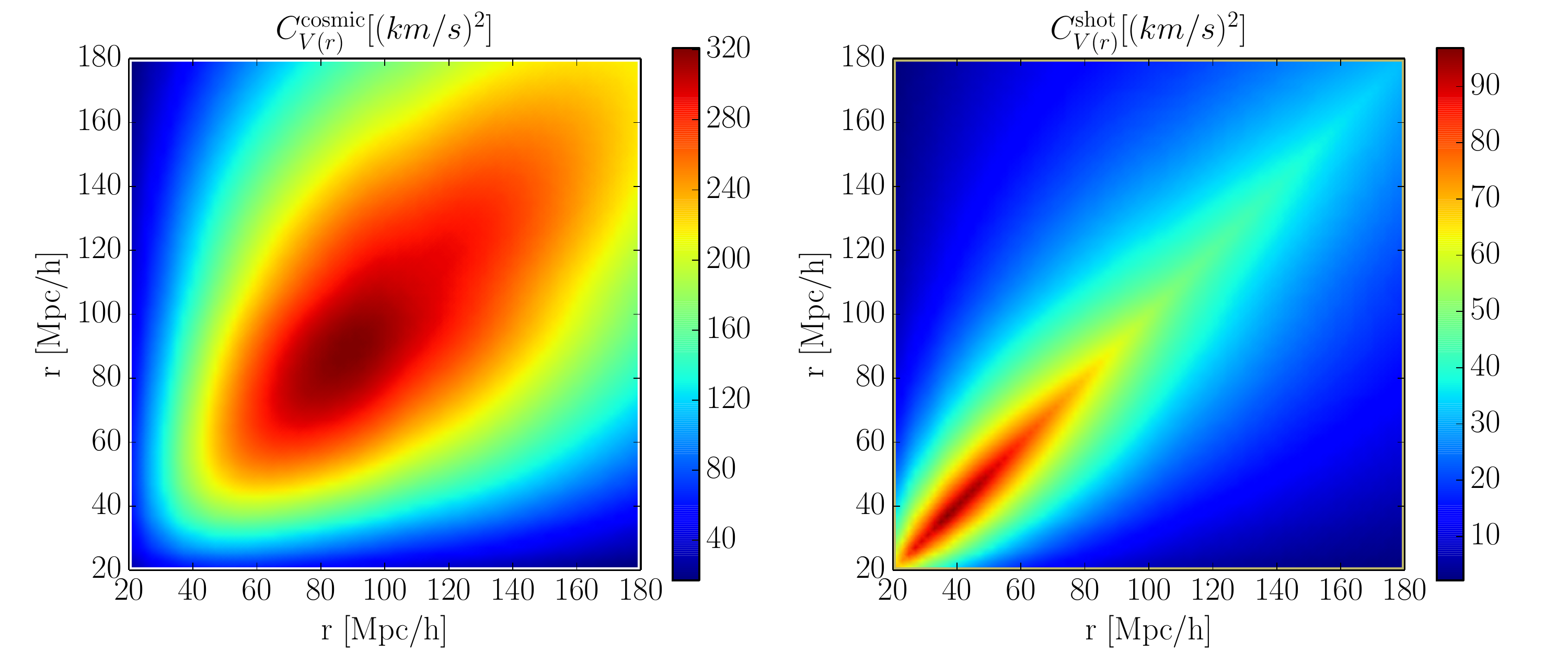}

\caption{ 2D contour plots of the cosmic variance [left panel] and the shot noise term [right panel] at redshift $0.1<z<0.2$ assuming a separation bin size of $\Delta r=2 \ \mathrm{Mpc/h}$, a lower mass limit of $M_{\mathrm{min}}=1\times10^{14} M_{\odot}$, and a sky coverage of $6000$ square degrees. Note that both terms have notable non-zero off-diagonal terms that affect the total inverse covariance used in the Fisher analysis, and that while the cosmic variance values are larger, the Gaussian noise term should not be neglected as it can have a significant effect, particularly for small separations.}
\label{fig:MeshCov}
\ec
\end{figure*}
%==================================================================

We include a contribution to the covariance due to the uncertainty in measuring the velocity given by  \cite{Bhattacharya:2007sk},
\bea
C_{\Vij}^{\mathrm{measurement}}(r,r')&=&\frac{2 \sigma_v^2}{N_{\mathrm{pair}}} \delta_{r,r'}
\eea
where $\sigma_v$ is the measurement error 
discussed in more detail in section \ref{sec:sigv}, and $N_{\mathrm{pair}}$ is the number of pairs in each separation bin given by
\bea
N_{\mathrm{pair}}(r,a)&=&\frac{n(a) V_s(a)}{2} \nonumber
\\
&&\times \left( V_{\Delta}(r)n(a) + 4 \pi r^2n(a) \xi_{h}(r,a) \Delta r \right). \hspace{0.5cm}
\eea
As shown in Figure \ref{fig:npair}, the number of cluster pairs increases rapidly with decreased minimum mass.
The measurement error term in the covaraince is proportional to $1/N_{\mathrm{pairs}}$ and will increase quickly with an increasing number of bins since the number of cluster pairs directly depends on the size of the $r$-bin.

Figure \ref{fig:CovBin} shows the diagonal elements of the  covariance matrix for the different covariance components, 
for a bin width of $\Delta r=2 \ \mathrm{Mpc/h}$. As  cluster separation increases, the covariance becomes dominated by cosmic variance, while at smaller  separations $\lesssim 40\  \mathrm{Mpc/h}$, the contributions from each of the terms becomes comparable. 
As a result of the multiple contributions to the covariance matrix, and their respective sensitivities to bin size and cluster separation, the total covariance matrix 
slightly depends upon the number of bins. The measurement error and shot noise can be reduced by choosing a coarser binning with the trade-off of decreased resolution and loss of information. On the other hand, the fractional contribution of the cosmic variance will increase as the size of the bins increases. Once the cosmic variance dominates nothing can be gained from a coarser binning. 
A very coarse binning marginally reduces the constraints, e.g. using $\Delta r = 20 \ \mathrm{Mpc/h}$ lowers the FoM by $30\%$ compared to $\Delta r = 2 \  \mathrm{Mpc/h}$, however, any bin size smaller than $\Delta r= 5 \  \mathrm{Mpc/h}$ leads to equivalent results. Throughout the analysis we assume a bin size of $\Delta r= 2 \  \mathrm{Mpc/h}$.

Off-diagonal covariances between cluster pairs of different separations are important.
Figure \ref{fig:MeshCov} shows the covariance contributions from cosmic variance and shot noise and indicates the comparative importance of off-diagonal terms. The off-diagonal contributions have a notable effect on the Fisher matrix 
amplitudes as a function of separation, giving rise to the differences between the left and right panels in Figure \ref{fig:CovBin}. The right panel shows the  effect on the Fisher matrix of changing key model assumptions, the minimum detectable cluster mass and the mean pairwise  velocity uncertainty.  Altering the mass limit has a larger effect than comparable changes to the the measurement error because the number of clusters and cluster pairs strongly depend on the limiting mass (see Figure \ref{fig:npair}), changing shot noise as well as the measurement error contribution to the covariance significantly.

%%%%%%%%%%%%%%%%%%%%%%%%%%%%%%%%%%%%%%%%%%%%
\subsection{Cosmological Model}
\label{sec:cosmomodel}
%%%%%%%%%%%%%%%%%%%%%%%%%%%%%%%%%%%%%%%%%%%%
For our analysis we consider constraints on 9 cosmological parameters:
\bea
\bold{p}&=&\{\Omega_bh^2, \Omega_mh^2,\Omega_k,\Omega_{\Lambda},w_0,w_a,
n_s,\ln A_s,\gamma\} \hspace{0.5cm}\label{parlist}
\eea
where $\Omega_b$, $\Omega_m$, $\Omega_k$ and $\Omega_{\Lambda}$ are  the dimensionless baryon, matter, curvature and dark energy densities respectively, $h$ is the Hubble constant in units of 100 km/s/Mpc, $w_0$ and $w_a$ are the dark energy equation of state parameters, such that the equation of state is $w(a)=w_0+(1-a)w_a$, $\gamma$ is the growth rate exponent, such that $f_g=\Omega_m(a)^\gamma$, and $n_s$ and $A_s$ are the spectral index and normalization of the primordial spectrum of curvature perturbations.

Throughout this paper we assume a fiducial model that is a $\Lambda$CDM cosmological model with  parameters consistent with those adopted in \cite{Laureijs:2011gra}: 
$ \Omega_bh^2=0.021805, \ \Omega_mh^2=0.1225,\ \Omega_k=0,\ \Omega_{\Lambda}=0.75,\ w_0=-0.95,\ w_a=0,\ n_s=1,\ \ln(10^{10}A_s)=3.1954$.

We calculate constraints on cosmological parameters using the Fisher Matrix formalism. The covariance between two parameters $p_\mu$ and $p_\nu$, from (\ref{parlist}), is given by
\bea
F_{\mu \nu}=\sum_{i}^{N_z}\sum_{p,q}^{N_r} \frac{\partial V(r_p,z_i)}{\partial p_{\mu}} Cov^{-1}_i(r_p,r_q,z_i) \frac{\partial V(r_p,z_i)}{\partial p_{\nu}},
\eea
where $Cov(r_p,r_q,z_i)$ is the covariance matrix between  two clusters pairs  as defined in \ref{sec:Cov}, including a redshift bin with mid-point $z_i$ and the clusters in each pair having comoving separations of $r_p$ and $r_q$.
$N_z$ and $N_r$ are the number of redshift and spatial separation bins, respectively.

We quote results in terms of the Dark Energy Figures of 
Merit (FoM) \cite{Albrecht:2006um} defined as
\bea
\mathrm{FoM}&=&\det\left[(F^{-1})\right]_{w_0,w_a}^{-1/2}
\\
 \mathrm{FoM}_{\mathrm{GR}}&=&\det\left[(F^{-1}_{\mathrm{GR}})\right]_{w_0,w_a}^{-1/2}.
\eea
$(F^{-1}_{\mathrm{GR}})_{w_0,w_a}$ is the $2\times2$ submatrix of the inverted Fisher matrix excluding the modified gravity parameter $\gamma$. This procedure is equivalent to marginalizing over the $7$ parameters (for MG) or $6$ parameters (for GR) of the model considered.

Throughout this paper we consider results in combination with either simply a Planck-like CMB prior or a Dark Energy Taskforce (DETF) \cite{Albrecht:2006um} prior that includes CMB, SN, and non-kSZ related LSS constraints on the background cosmological and dark energy parameters. We do not include a prior on the modified gravity parameters unless stated otherwise.
For the  Planck-like CMB survey, we consider complementary constraints on the cosmological parameters from the temperature ($T$) and polarization ($E$) measurements up to $l=3000$ as summarized in Table \ref{tab:Planck_spec} \footnote{www.rssd.esa.int/SA/P LANCK/docs/Bluebook-ESA-SCI(2005)1V2.pdf}.

%=================================================================
\begin{table}
\begin{center}
\begin{tabular}{|c |c| c| c|}
        \cline{2-4}
\multicolumn{1}{c}{} 
& \multicolumn{3}{|c|}{Frequency (GHz)}\\
        \cline{2-4}
\multicolumn{1}{c|}{} & 100 & 143 & 217\\
\hline
\multicolumn{1}{|c|}{$f_{\mathrm{sky}}$} & \multicolumn{3}{|c|} {0.8}\\
\hline
\multicolumn{1}{|c|}{$\theta_{\mathrm{FWHM}}$(arcmin)} & 10.7 & 8.0 & 5.5\\
\hline
\multicolumn{1}{|c|}{$\sigma_{T}$($\mu$K)} & 5.4 & 6.0 & 13.1\\
\hline
\multicolumn{1}{|c|}{$\sigma_{E}$($\mu$K)} & - & 11.4 & 26.7\\
\hline
\end{tabular}
\caption{CMB survey specifications, for the sky coverage, $f_{\mathrm{sky}}$, beam size, $\theta_{\mathrm{FWHM}}$, and noise levels per pixel for the temperature and polarization detections at 3 frequencies, for a Planck-like survey. }
\label{tab:Planck_spec}
\end{center}
\end{table}
%------------------------------------------------------------------------------------------------------------------

%%%%%%%%%%%%%%%%%%%%%%%%%%%%%%%%%%%%%%%%%%%%
\subsection{Survey Specifications}
\label{sec:survey_spec}
%%%%%%%%%%%%%%%%%%%%%%%%%%%%%%%%%%%%%%%%%%%%
We forecast cosmological constraints  for three different combinations of surveys: 1) a current (Stage II) 
CMB survey, such as ACTPol  \cite{Niemack:2010wz}, combined with a galaxy sample that includes spectroscopic redshifts, such as SDSS BOSS \cite{Comparat:2012hz}, 2) a near-term (Stage III) 
 survey, such as Advanced ACTPol \cite{Calabrese:2014}, also combined with SDSS BOSS, and 3) a longer-term (Stage IV) 
 survey, such as CMB-S4 \cite{CMBS4neu:2013}, combined with a next generation spectroscopic survey, such as DESI \cite{Levi:2013gra}.
 
The mean cluster pairwise velocity can be measured by cross-correlating the kSZ signal with cluster positions and redshifts. For the cluster sample, we assume that a spectroscopic survey provides redshifts to luminous red galaxies (LRGs) over an overlapping area with the CMB survey.    
Recent studies show that the kSZ signal can be extracted from the CMB maps using LRGs of the BOSS survey as a proxy for clusters \cite{2011ApJ...736...39H}.  Using LRGs creates a large, precise positioned sample of tracers to extract the kSZ correlation.

%=================================================================
\begin{table}[!t]
\begin{center}
	\begin{tabular}{|l|l|c|c|c|}
	\cline{3-5}
	 \multicolumn{2}{c}{} & \multicolumn{3}{|c|}{ Survey Stage }
\\	\hline
 Survey & Parameters \  & \ II \ & \ III \ & \  IV \    \\
    \hline
 CMB&    $\Delta T_{\mathrm{instr}}$ ($\mu K \mathrm{arcmin}$) \ & 20 &  7  & 1
    \\ \hline
 \multirow{4}{*}{Galaxy} \ &   $z_{\mathrm{min}}$ &0.1&0.1&0.1
    \\
   &  $z_{\mathrm{max}}$  \ &0.4&0.4&0.6
    \\
& No. of $z$ bins, $N_z$   \   &3&3&5
    \\
  &  $M_{\mathrm{min}}$ ($10^{14}M_\odot $) \ & $1 $ & $1$ &  $0.6$
  \\  \hline
  Overlap &    Area (1000 sq. deg.) \ & 4 & 6& 10
    \\ \hline
    \end{tabular}
    \caption{Reference survey specifications used to model Stage II, III and IV  
    kSZ cluster surveys. The expected instrument sensitivity of the CMB survey, $\Delta T_{\mathrm{inst}}$, along with the assumed optical large scale structure survey redshift range $z_{\mathrm{min}}<z<z_{\mathrm{max}}$,  redshift binning, and minimum detectable cluster mass, $M_{\mathrm{min}}$ are shown. We consider an effective sky coverage 
    by estimating the degree of overlap between the respective CMB and optically selected cluster datasets.}
    \label{tab:kSZ_surveys}
\end{center}
\end{table}
%=================================================================

However, there are several factors that need to be considered in using LRGs as cluster tracers.
LRGs are not perfect tracers of a cluster's center, with perhaps 40\% of bright LRGs and 70\% of faint LRGs off-centered, satellite galaxies \cite{Hikage:2012zk} that may be related to cluster mergers \cite{Martel:2014zoa}.
The imprecise match between LRGs and clusters could lead to detrimental misalignments, such as trying to extract the kSZ signal from positions that are not associated with clusters or an incomplete cluster catalog if spectroscopic measurements of an LRG near the cluster center were not obtained.
The theoretical mean cluster pairwise velocity is an observable averaged over all cluster pairs assuming a complete sample above a limiting minimum mass. While \citet{2011ApJ...736...39H} optimize the angular size of the CMB sub-map used in the stacking approach to minimize the overall covariance, this does not ensure that the cluster sample obtained from the LRGs is complete. Further studies are needed to quantify the effects of using LRGs as cluster tracers and ensure that no bias is introduced in the analysis before this approach can be used for cosmological constraints.
Another issue is that the uncertainty in the minimum mass of the cluster sample associated with the LRGs is difficult to estimate, although, the minimum mass uncertainty could be treated as an additional nuisance parameter in the analysis.

To acknowledge these issues in our forecasts, we assume a  
scenario that aims to maximize cluster completeness and purity, with a well defined cluster mass cut-off, rather than cluster number density.  We select a survey area that has photometric and spectroscopic galaxy catalogs and overlapping CMB kSZ data. Specifically, we consider BOSS and a DESI-like survey, 
for which we expect 
photometric catalogs to exist over the survey area. We note that Euclid spectroscopic and imaging surveys, and LSST imaging with overlapping WFIRST imaging and spectroscopy would also provide future valuable datasets at higher redshifts. The uncertainties in the cosmological parameters 
evolve as the square root of the sky coverage. Requiring spectroscopic redshifts, e.g. from BOSS, limits the survey area, but provides confidence that the comoving cluster separation can be accurately calculated as in \cite{Hand:2012ui}. Photometric information allows cluster detection, and mass estimates, using  algorithms, such as the friends-of-friends, as used in redMaPPer \cite{Rykoff:2013ovv},  to maximize the completeness and purity of the cluster sample, with the drawback of a limited number of clusters and a volume-limited catalog. A study of using only photometric information to extract the kSZ signal can be found in \cite{Keisler:2012eg}.

The survey specification assumed in our analysis for the CMB and large scale structure Stage II, III and IV surveys are given in Table \ref{tab:kSZ_surveys}. We assume a BOSS-like spectroscopic survey for Stage II and III and a DESI-like Stage IV survey with redshift ranges that are determined by the redshift coverage of the LRG sample and assume joint photometric survey data. We assume Stage II and Stage III have access to the same or comparable LRG surveys so retain the same limiting mass, but do slightly increase CMB overlap with these data due to the larger survey area planned for Advanced ACTPol \cite{Calabrese:2014}. For Stage IV we assume a deeper LRG survey that provides lower minimum mass, higher $z$, and larger overlap.
Our minimum mass assumptions are conservative, and will likely be improved upon at each respective stage. As an example, the LSST  survey projects that the minimum detectable cluster mass at $z\sim 0.6$ will be lower than $\sim 5\times 10^{13}M_{\odot}$ after a single visit image in all bands, and be better than $10^{13}M_{\odot}$ in all bands in the complete ten-year survey \cite{Abell:2009aa}. Similarly the SDSS-derived MaxBCG Catalog  already achieves 90\% purity and  $>$85\% completeness for clusters of masses exceeding $10^{14}M_{\odot}$ \cite{Koester:2007bg}.

 %=================================================================
\begin{table}[!t]
\begin{center}
    \begin{tabular}{|l|l|c|c|c|c|c|c|}
        \cline{3-7}
\multicolumn{2}{c|}{} & \multicolumn{5}{|c|}{ Redshift bin} \\
 \cline{3-7}
  %RACH   z
\multicolumn{2}{c|}{} & \ $0.15$ \ & \ $0.25$ \  & \ $0.35$ \ & \ $0.45$  \ &  \ $0.55$ \
    \\
    \hline
     \multicolumn{2}{|l|}{$10^3 \tau$} 
  &3.45  &2.27&  1.84 & 1.45&  1.20
    \\
    \hline
     \multicolumn{2}{|l|}{$(\Delta_\tau / \bar{\tau})^2$} 
    &\multicolumn{5}{|c|}{ 0.15}
    \\
    \hline
  \multicolumn{2}{|l|}{$\sigma_{\tau}$ (km/s)}&\multicolumn{5}{|c|}{ 120}
     \\
    \hline
  $\sigma_{\mathrm{instr}}$ & Stage II& 290 &  440 &   540&-&-
    \\
      \cline{2-7}
(km/s)    & Stage III &100&  150 & 190 &-&-
    \\
    \cline{2-7}
& Stage IV& 15&  22& 27&  34&  42

    \\
    \hline
$\sigma_{v}$ & Stage II&310&  460 & 560 &-&-

    \\
      \cline{2-7}
   (km/s)     &Stage III&160& 200&   230 &-&-
    \\
      \cline{2-7}
        & Stage IV&120&  120& 120& 120&  130
    \\
    \hline
    \end{tabular}
    \caption{The assumed  individual contribution from instrument sensitivity, $\sigma_{\mathrm{instr}}$, and uncertainty in $\tau$, $\sigma_{\tau}$. The values of $\tau$ and fractional uncertainty in $\tau$, $(\Delta\tau / \bar{\tau})^2$, are estimated from simulations assuming a convolution over a 1.3$'$ beam. $\sigma_v$ is the total measurement uncertainty for the reference case. }
  \label{tab:tau}
\end{center}
\end{table} 
%=================================================================

The measurement error for the radial peculiar velocity, $v$, of a cluster 
is a combination of the instrumental sensitivity as well as the uncertainty in the optical depth, $\tau$, for each cluster as the kSZ signal is proportional to $\tau$ as follows \cite{Sunyaev:1980nv},
\bea
\frac{\Delta T_{\mathrm{kSZ}}}{T_{\mathrm{CMB}}}= -\frac{v}{c} \tau, \label{eq:tksz}
\eea 
where $T_{\mathrm{CMB}}$ is the temperature of the CMB. We estimate the total measurement error by adding those two sources of uncertainty in quadrature as
\bea
\sigma_v=\sqrt{ \sigma_{\mathrm{instr}}^2 +\sigma_{\tau}^2}  \label{eq:sigv}.
\eea
The accuracy of the instrument is given by
\bea
\sigma_{\mathrm{instr}}= \frac{\Delta T_{\mathrm{instr}}}{\Delta T_{\mathrm{kSZ}}}\times v=\frac{\Delta T_{\mathrm{pixel}}/\sqrt{N_{\mathrm{pixel}}}}{\tau v/c T_{\mathrm{CMB}}} \times v
 \eea
  where $\Delta T_{\mathrm{pixel}}$ is the sensitivity of the instrument per pixel and $N_{\mathrm{pixel}}$ being the number of pixels of a cluster.
We assume that an average size cluster will have $N_{\mathrm{pixel}}\approx 4$ and an instrument sensitivity as summarized in Table \ref{tab:kSZ_surveys}. The uncertainty in the optical depth   is given by
  \bea
   \sigma_{\tau}=\frac{\Delta \tau}{\tau} \times v.
   \eea

Assumed uncertainties contributing to the measurement error are summarized in Table \ref{tab:tau}.  We use the scatter in the optical depth, $|\Delta\tau/\bar\tau|$, and the mean value of $\tau$ from simulations \cite{Battaglia:2014}  \footnote{N. Battaglia: private communication},
 to obtain an indicative 
  estimate for the intrinsic dispersion in $\tau$ averaged over all cluster masses. For the fiducial analysis we do not include any further dispersion arising from potential additional measurement accuracy in determining $\tau$.  Section \ref{sec:measure_error} includes a discussion of the impact of additional factors affecting the measurement error on the cosmological constraints.

%%%%%%%%%%%%%%%%%%%%%%%%%%%%%%%%%%%%%%%%%%%%
\section{Analysis }
\label{sec:analysis}
%%%%%%%%%%%%%%%%%%%%%%%%%%%%%%%%%%%%%%%%%%%%
Section \ref{sec:forecast_summary} summarizes and compares the results of each survey. The effect of modeling assumptions on the minimum detectable cluster mass, the minimum cluster separation considered, the measurement error, and the dark energy model are discussed in sections \ref{sec:Mmin}-\ref{sec:deparam}.

%=================================================================
\begin{figure}[!t]
\bc
{\includegraphics[width=0.48\textwidth]{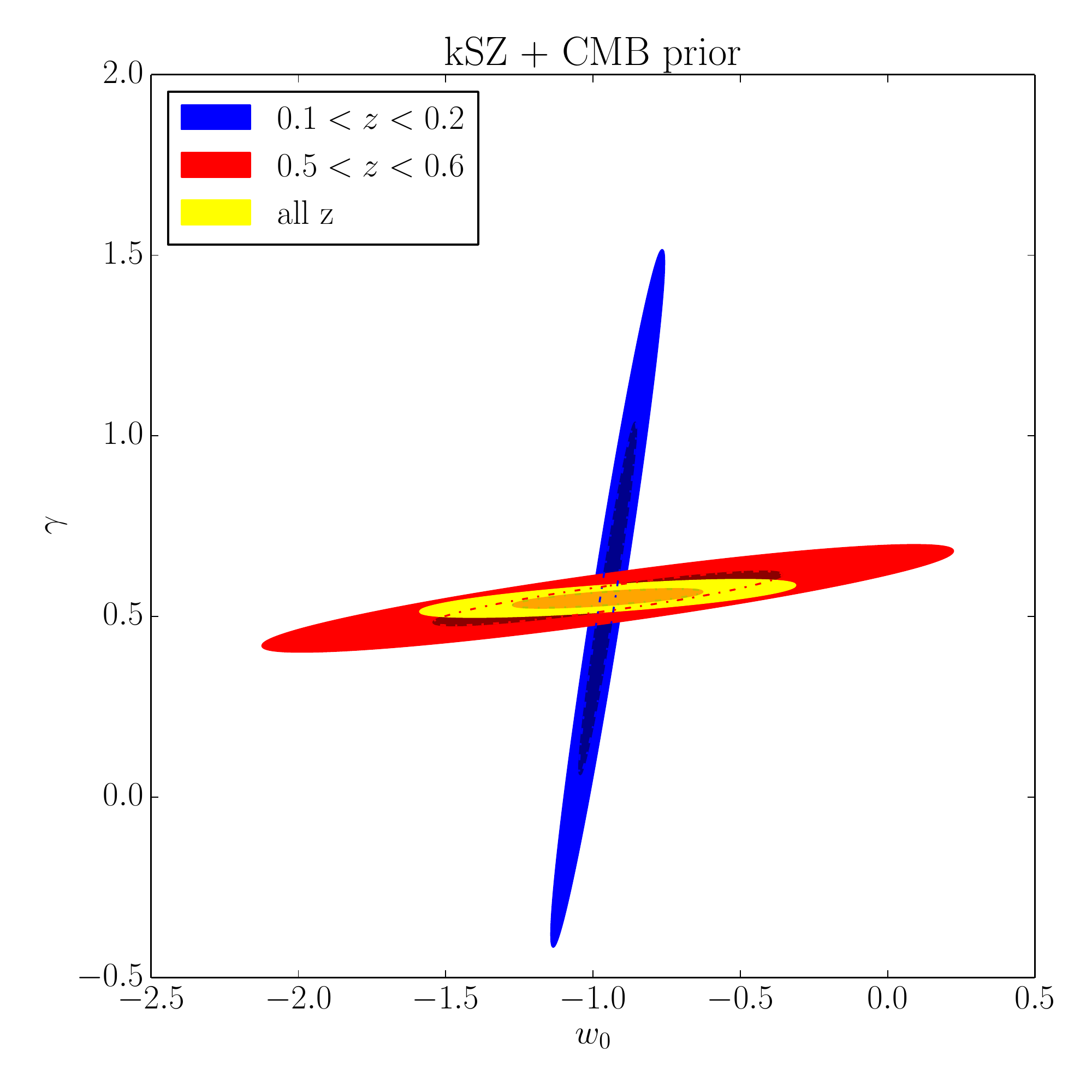}
}
\caption{2D projected likelihoods for the  $w_0-\gamma$ parameter space, showing the $68\%$ and 95\% confidence levels for Stage IV-like survey are shown for two well-separated spectroscopic redshift bins, $0.1<z<0.2$ (blue) and  $0.5<z<0.6$ (red), and when all five redshift bins, $0.1<z<0.6$ (yellow) are considered when combined with Planck-like CMB priors. The inclusion of multiple redshift bins breaks degeneracies between $w$ and $\gamma$ and improves the kSZ driven constraints on the growth history.}
\label{fig:Ellipse_zbins}
\ec
\end{figure}
%=================================================================

%=================================================================
\begin{figure}[!t]
\bc
{\includegraphics[width=0.48\textwidth]{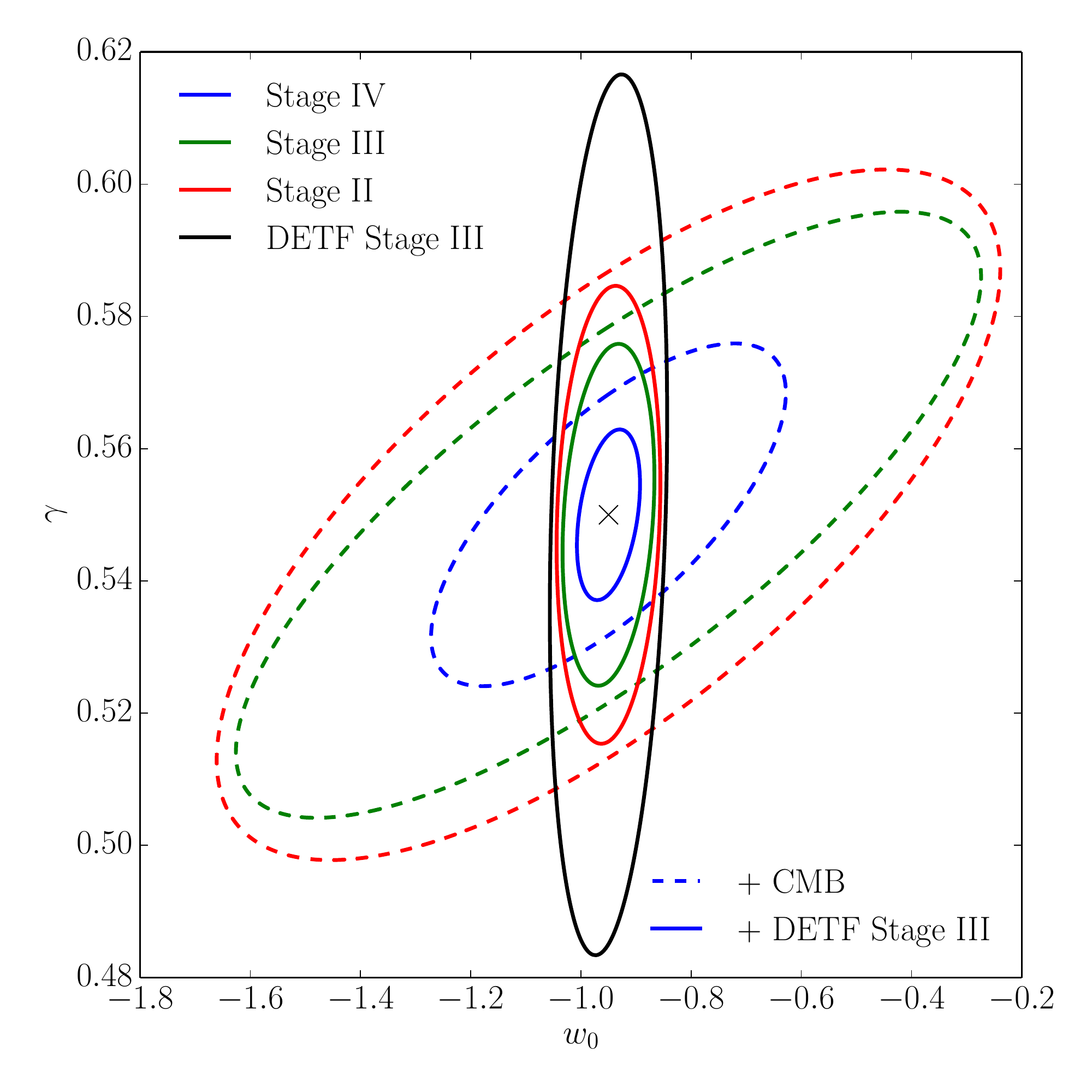}
}
\caption{2D projected likelihoods for   $w_0-\gamma$ parameter space, showing the $68\%$ confidence levels for Stage II (red), III (green) and IV (blue)-like surveys when combined with Planck-like CMB priors only (dashed) and  DETF stage III GR priors (solid, excluding DETF constraints on $\gamma$) \cite{Albrecht:2006um}. For comparison, the projected  DETF Stage III constraints alone (including $\gamma$), that includes CMB, SN and non-kSZ related LSS constraints, are shown (black solid line).}
\label{fig:Ellipse}
\ec
\end{figure}
%=================================================================

%%%%%%%%%%%%%%%%%%%%%%%%%%%%%%%%%%%%%%%%%%%%
\subsection{Potential kSZ constraints on dark energy and modified gravity}
\label{sec:forecast_summary}
%%%%%%%%%%%%%%%%%%%%%%%%%%%%%%%%%%%%%%%%%%%%

In this section we discuss the potential of upcoming kSZ surveys  to constrain dark energy and modified gravity parameters. 
 Figure \ref{fig:Vij} shows  both the equation of state, driving the expansion history, and $\gamma$, that modifies the growth history of density perturbations, have qualitatively similar effects on the pairwise velocity function through their effect on the linear growth factor. For cluster measurements in each individual redshift bin this creates a degeneracy between the equation of state and $\gamma$ parameters. 
As shown in Figure \ref{fig:Ellipse_zbins}, the use of multiple redshift bins allows the differences in the evolution of the growth rate for the dark energy and modified gravity parameters to be distinguished. The constraints on $w_0$ and $\gamma$ from the low and high redshift bins are markedly orthogonal;  in combination this complementarity tightens the constraints, in particular on the growth factor. In Figure \ref{fig:Ellipse} we present the 2D marginalized constraints  in the $w_0-\gamma$ parameter plane for kSZ in combination with a Planck-like CMB and DETF priors on all parameters excluding $\gamma$, including CMB, BAO, weak lensing and supernovae measurements for a combination of Stage III like surveys \cite{Albrecht:2006um}. The suite of Stage III DETF-motivated observables would provide stronger constraints on the equation of state, through the addition of geometric measurements that constrain the expansion history. These break the degeneracy between $w_0$ and $\gamma$ from kSZ and CMB measurements alone. With the addition of a CMB prior on the data, however, the data can constrain $\gamma$ to 10\%, 8\% and 5\% respective in the Stage II through IV survey specifications.  The kSZ is
a less powerful tool for constraining the dark energy equation of state. A Stage IV-like survey can achieve figure of merits of $\mathrm{FoM}_{\mathrm{GR}}=61$ with a CMB prior (which has $\mathrm{FoM}_{\mathrm{GR}}=1.15$ alone), and $\mathrm{FoM}=292$ with DETF Stage III data included ($\mathrm{FoM}_{\mathrm{GR}}=116$).

Complementary, contemporaneous constraints from baryonic acoustic oscillations and type Ia supernovae will provide significantly tighter constraints on the background expansion history and the equation of state. If we include the impact of a DETF Stage III prior on all parameters, excluding the growth factor, the degeneracy between the equation of state and growth factor is significantly reduced and the projected constraints on $\gamma$ are improved, with fractional errors of 5\% and 2\% for Stage III and Stage IV surveys.
Table \ref{tab:results} summarizes the dark energy figure of merit (FoM), assuming modified gravity (9 parameters, marginalizing over $\gamma$), General Relativity (GR) (8 parameters, fixing $\gamma$) and a flat, General Relativity cosmology  
(7 parameters, fixing $\gamma$ and $\Omega_k$), as well as the $1\sigma$
 constraints of $w_0$, $w_a$ (marginalizing over $\gamma$)and $\gamma$   for a Stage II, Stage III and Stage IV like survey, as specified in Table \ref{tab:kSZ_surveys}.

%=================================================================
\begin{table*}[!htb]
\begin{center}
\begin{tabular}{|l|l|r|r|r|| r|r|r|| r|r|r||}
\cline{3-11}
\multicolumn{2}{c}{} & \multicolumn{3}{|c||}{Fiducial assumptions}&\multicolumn{3}{|c||}{+ Uncertainty in  $M_{\mathrm{min}}$}&\multicolumn{3}{|c||}{+ Lower $M_{\mathrm{min}}$}\\
\cline{3-11}
\multicolumn{2}{c|}{}   & Stage II & Stage III & Stage IV & Stage II &   Stage III &  Stage IV& Stage II &   Stage III &  Stage IV\\
\hline 
\multirow{5}{*}{+CMB} & $\mathrm{FoM}_{MG}$            &      6  &       9   &     33                             &            4   &              6   &            27  &      8   &      12  &     81                      \\  
& $\mathrm{FoM}_{\mathrm{GR}}$            &      8     &      14  &     61                           &            6   &              9   &            43 &     12   &      20   &    110                             \\
& $\mathrm{FoM}_{\mathrm{flat}}$          &     37   &      57     &    128               &   29   &              39   &            94          &     52   &      68  &    206                   \\
\cline{2-11}
 &$\sigma(w_0)$         &      0.72 &       0.68 &      0.33    &    0.73   &              0.69   &            0.33             &      0.63 &       0.55 &      0.18               \\
& $\sigma(w_a)$         &      2.6 &       2.5 &      1.2           &     2.6   &              2.5   &            1.2            &      2.3 &       2.0 &      0.6 \\
& $\Delta\gamma/\gamma$ &    0.10 &    0.08 &    0.05     &  0.10   &              0.09   &            0.05            &      0.07 &       0.06 &      0.02          \\ \hline
\multirow{5}{*}{+DETF} & $\mathrm{FoM}_{MG}$            &  131   &       152    &       273  &  \multicolumn{6}{c}{}    \\  
& $\mathrm{FoM}_{\mathrm{GR}}$            &      133    &       156 &      292        & \multicolumn{6}{c}{}       \\
& $\mathrm{FoM}_{\mathrm{flat}}$          &      181   &       213     &     405     & \multicolumn{6}{c}{}              \\
\cline{2-5}
 &$\sigma(w_0)$         &   0.10 & 0.08 & 0.06 &   \multicolumn{6}{c}{}           \\ 
& $\sigma(w_a)$         &      0.29&       0.26 &      0.21             & \multicolumn{6}{c}{}         \\
& $\Delta\gamma/\gamma$ &   0.06 &    0.05 &    0.02       & \multicolumn{6}{c}{}         \\
\cline{1-5}
\end{tabular}
\caption{[Left columns] Results for the reference survey assumptions as summarized in Table \ref{tab:kSZ_surveys} including [top rows] Planck priors and [lower rows]  constraints on the background cosmological parameters (excluding $\gamma$) from the DETF Stage III survey. For reference, the Planck-like Fisher matrix alone has $\mathrm{FoM}_{\mathrm{GR}}=1.15$ and DETF has $\mathrm{FoM}_{\mathrm{GR}}=116$. [Central columns] Results in which the impact of an uncertainty in the exact minimum mass of the cluster sample, $M_{\mathrm{min}}$, is included by marginalizing over $M_{\mathrm{min}}$ as a nuisance parameter with a $15\%$ prior imposed. [Right columns] Results for a more optimistic mass cut-off of $M_{\mathrm{min}}=4\times 10^{14} M_{\odot}$ for Stage II and III and  $M_{\mathrm{min}}=1\times10^{13} M_{\odot}$ for Stage IV with marginalization over $M_{\mathrm{min}}$ with a $15\%$ prior imposed as well. Constraints as a function of $M_{\mathrm{min}}$ are also shown at the top of Figure \ref{fig:sspecs}. }
\label{tab:results}
\end{center}
\end{table*}
%=================================================================
These results could provide valuable complementary constraints to those on $\gamma$ from spectroscopic galaxy clustering surveys. Projections include constraints of $\Delta\gamma/\gamma\simeq 5\%$ from measurements at $z>0.65$ using OII for a DESI-like survey \cite{Font-Ribera:2013rwa}, and comparable from a Euclid-like $H\alpha$ survey, for which  \cite{Amendola:2012ys} projected $\Delta \gamma/\gamma=4\%$ (assuming a luminosity function \cite{Geach:2009tm} that has since been revised downwards to lower $H\alpha$ number counts \cite{Wang:2012bx, Colbert:2013ita}).

%%%%%%%%%%%%%%%%%%%%%%%%%%%%%%%%%%%%%%%%%%%%
\subsection{Dependence on minimum mass of the galaxy cluster sample}
\label{sec:Mmin}
%%%%%%%%%%%%%%%%%%%%%%%%%%%%%%%%%%%%%%%%%%%%

For  cluster abundance measurements knowing the  precision with which the minimum mass is known is important. To assess the degree of precision required for the pairwise measurements we consider the impact on the cosmological constraints of marginalizing over the minimum mass, with a $15\%$ prior on $M_{\mathrm{min}}$. The middle panel of Table \ref{tab:results} shows the effects of this marginalization: the constraints are loosened only slightly compared to the fiducial case, that has no marginalization over the minimum mass. This implies that a precise knowledge of the minimum mass is not crucial to achieve cosmological constraints. 
An explanation for the comparative insensitivity of the dark energy constraints to uncertainties in $M_{min}$, can be understood with reference to Figure \ref{fig:Vij}. While varying  dark energy parameters and $M_{min}$ both change the large scale pairwise velocity amplitude the minimum mass also changes the shape of the pairwise velocity function. This means that uncertainties in the minimum mass can be discerned from those in dark energy, and do not translate into a comparable degradation of constraints on $w$ or $\gamma$.

\begin{figure*}[t]
\bc
{
\includegraphics[width=0.48\textwidth]{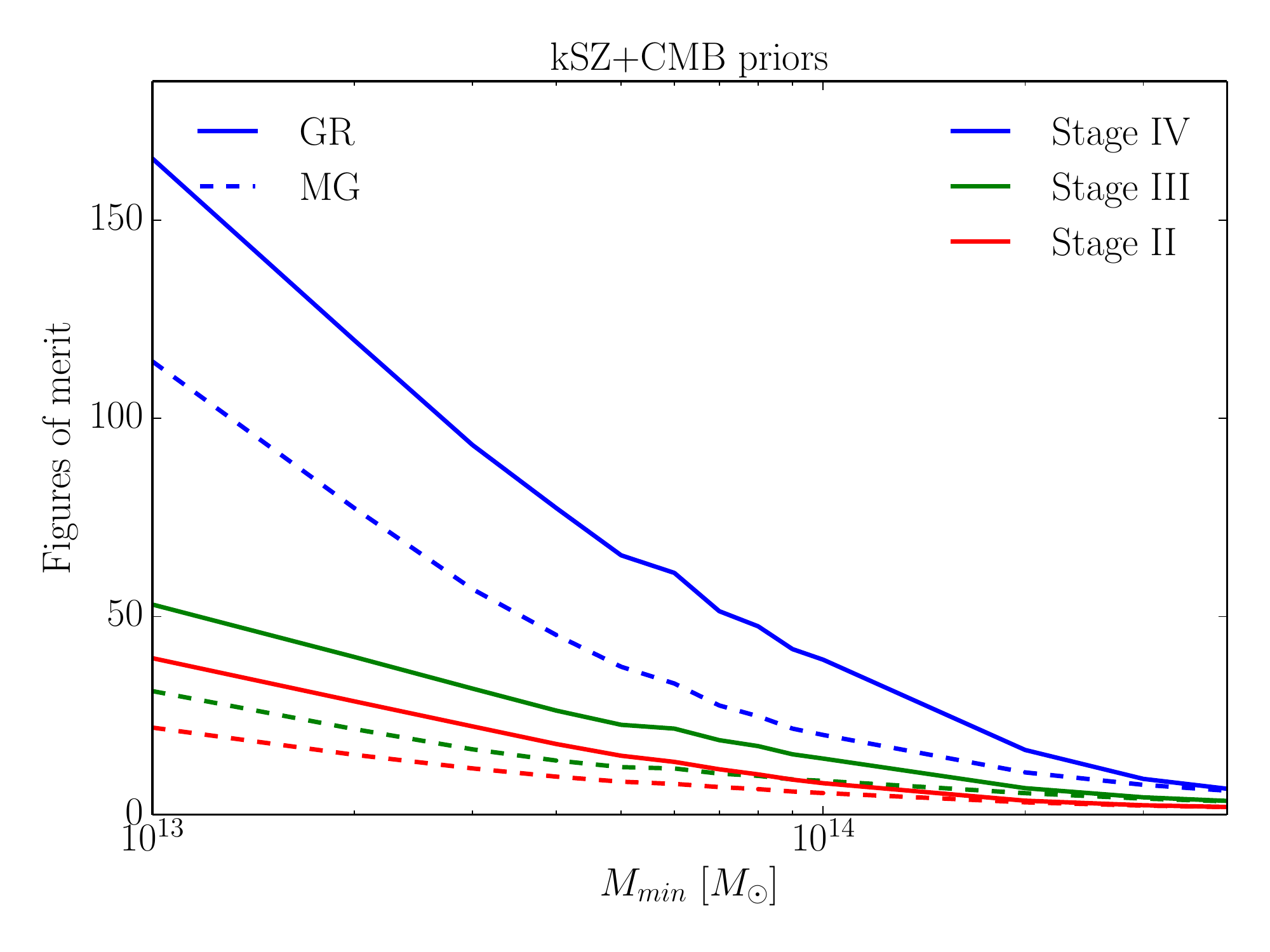} 
\includegraphics[width=0.48\textwidth]{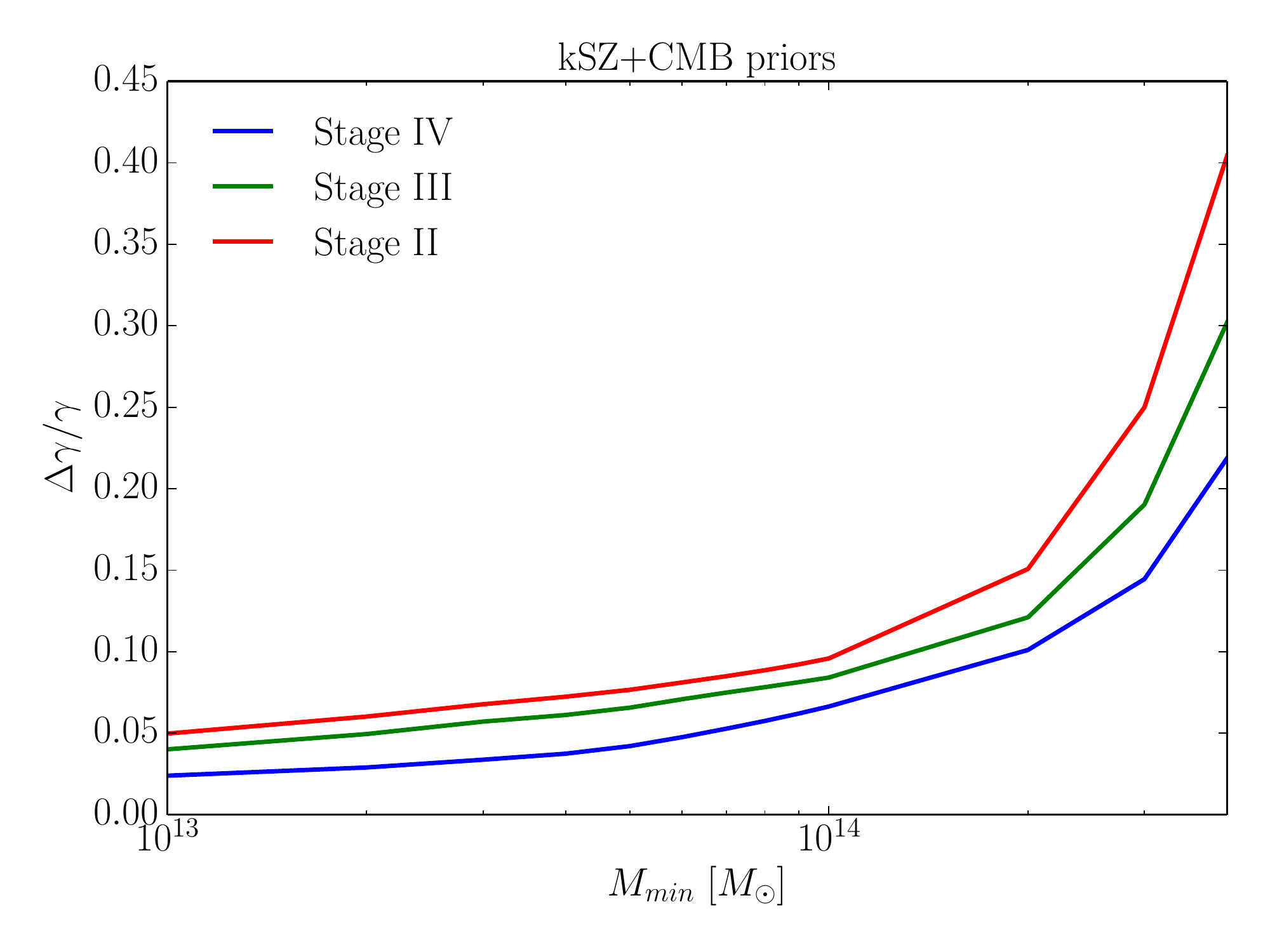} 
}
\\
{
\includegraphics[width=0.48\textwidth]{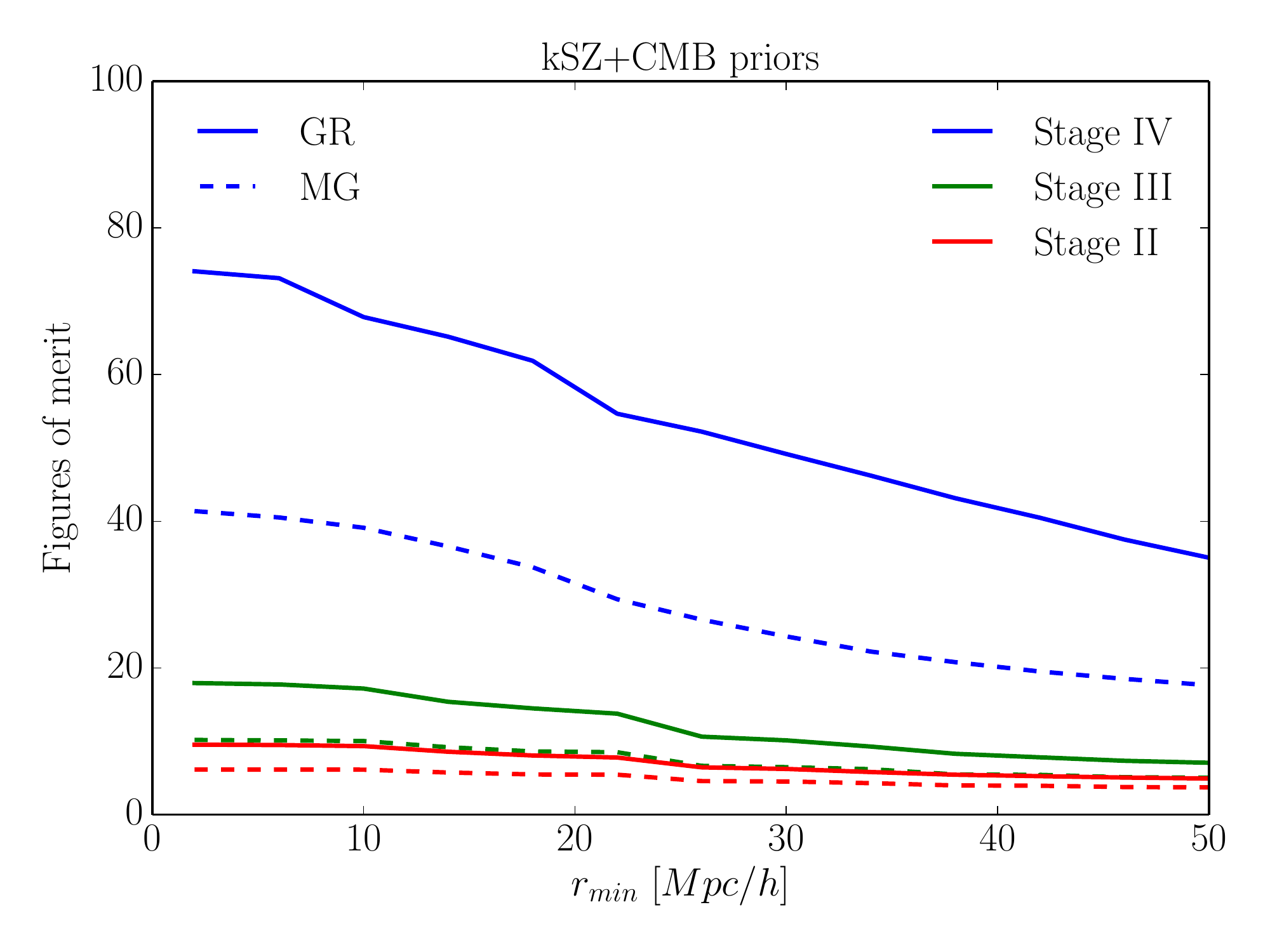} 
\includegraphics[width=0.48\textwidth]{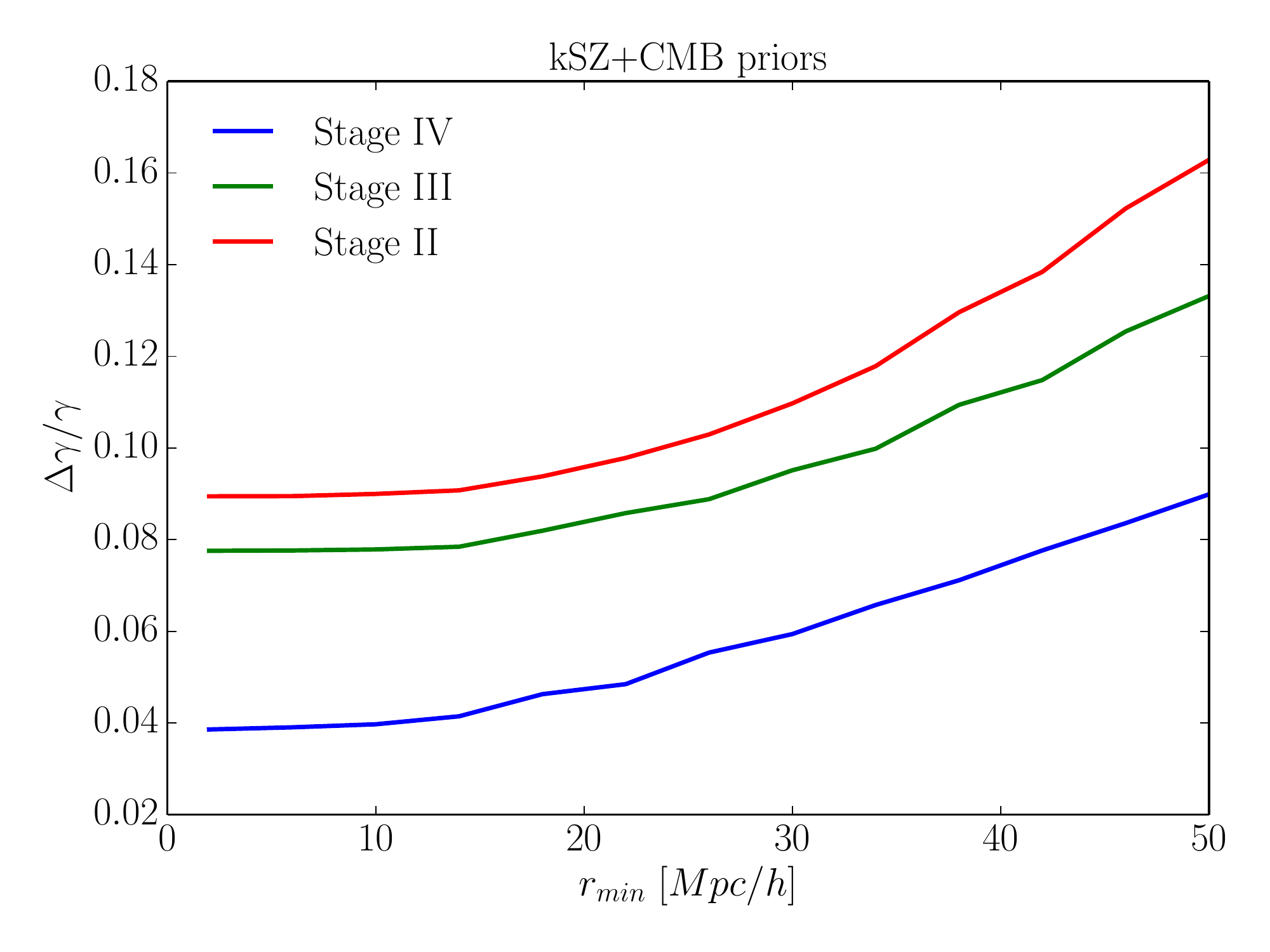}  
}
\caption{[Upper panel] The impact  of the assumed  minimum mass of the cluster sample, $M_{\mathrm{min}}$, on the dark energy figures of merit (FoM) and uncertainty on the growth factor, $\Delta \gamma/\gamma$, for the Stage II (red), Stage III (green) and Stage IV (blue) reference survey specifications (as given in Table \ref{tab:kSZ_surveys}) with a Planck-like CMB prior on all parameters except $\gamma$.  FoM plots show results assuming standard general relativity (`GR', solid lines) and when  the growth factor is marginalized over (`MG', dashed lines). [Lower panel] 
The impact of including observations on small scales, denoted by the minimum separation $r_{\mathrm{min}}$. While including smaller-scale observations below $\sim$20 Mpc$/h$ would appear to improve both  the FoM and $\Delta \gamma/\gamma$, as discussed in the text, we note that caution must be used in including these scales, with the potential for additional theoretical uncertainties, not included here, as non-linear effects become important.
}
\label{fig:sspecs}
\ec
\end{figure*}

The measurement uncertainty on the mean pairwise velocity decreases with the number of clusters used for the cross-correlation. The upper panels of Figure
\ref{fig:sspecs} presents the dependence of the FoM and $\Delta\gamma/\gamma$ constraints on the assumed minimum observed mass.
The  increased number density of clusters and cluster pairs arising from a lower mass bound, below $\sim10^{14}M_{\odot}$, significantly improves the statistical uncertainties in the pairwise velocity.
For our analysis we integrated over a Jenkins mass function using the minimum observed mass as our lower limit. As the number density of clusters drops off quickly for higher masses the constraints deteriorate strongly for a minimum mass above $M> 2\times 10^{14} M_{\odot}$.

%=================================================================
%=================================================================

Assuming that the 
complications discussed in section \ref{sec:survey_spec}, in determining LRG centrality and cluster mass estimates,  can be controlled, in principle one could achieve much higher number densities and a smaller minimum mass. This would increase the number of pairs in the cluster sample. The window functions for lower mass halos would include additional information  in the mass averaged statistics from the power spectrum at smaller scales that would lead to tighter constraints on the cosmological parameters. The right 
columns of Table \ref{tab:results} show the results assuming a more optimistic mass cut-off than the reference case, $M_{\mathrm{min}}=4\times10^{13} M_{\odot}$ for Stage II and III and $M_{\mathrm{min}}=1\times10^{13} M_{\odot}$ for Stage IV. To account for the uncertainty in mass we marginalize over the minimum mass assuming a $15\%$ prior. The GR figures of merit, with a CMB prior, are improved from $\mathrm{FoM}_{\mathrm{GR}}=8$ and $\mathrm{FoM}_{\mathrm{GR}}=14$ for Stage II and III to $\mathrm{FoM}_{\mathrm{GR}}=12$ and $\mathrm{FoM}_{\mathrm{GR}}=20$, compared to the reference scenarios, and by a factor of $1.8$ for Stage IV. The uncertainty in $\gamma$ reduces to $\Delta\gamma/\gamma=0.07$, $0.06$, and $0.02$ for Stage II, III and IV.

%=================================================================
\begin{figure*}[!tb]
\bc
{\includegraphics[width=0.48\textwidth]{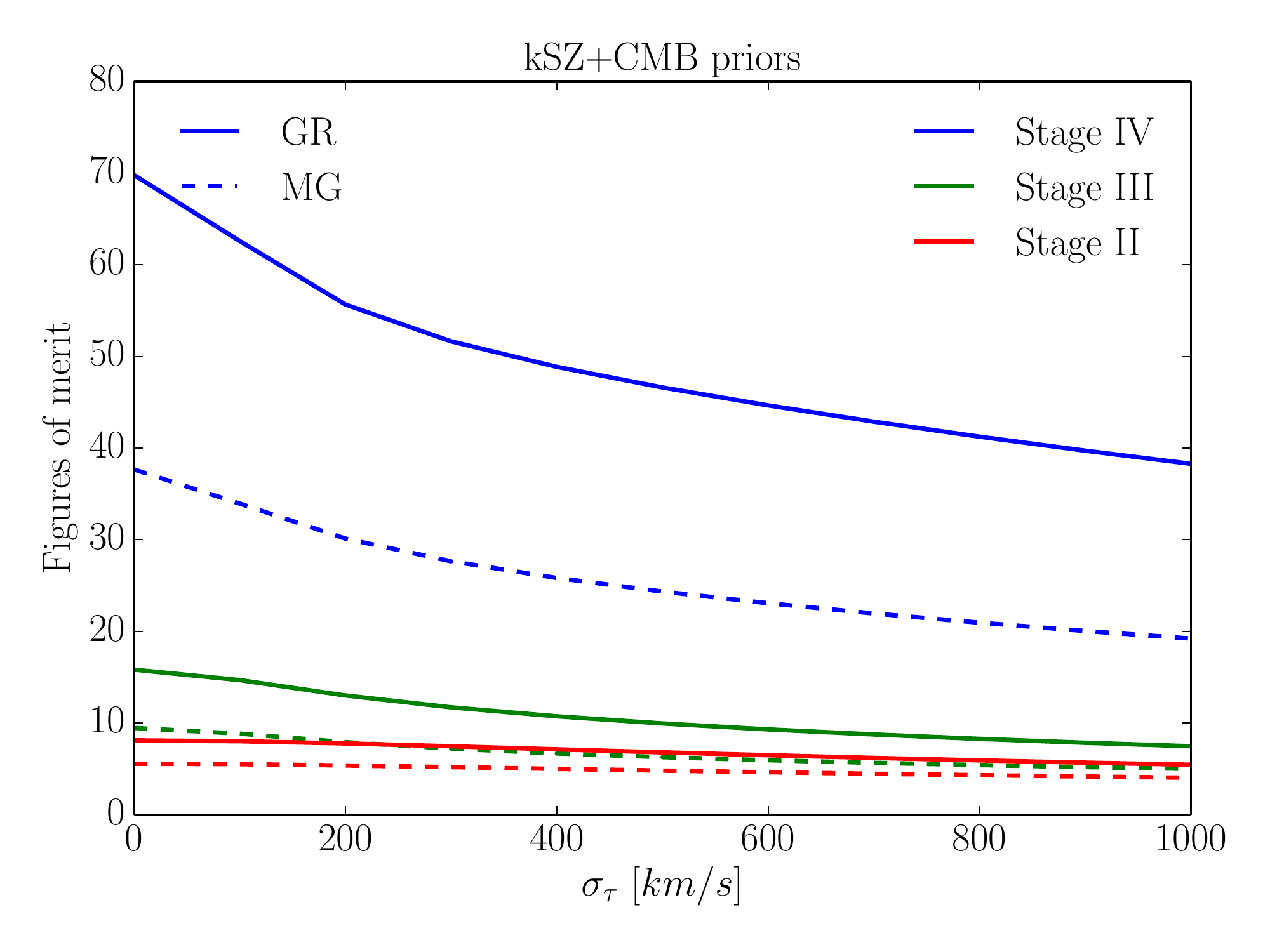} 
\includegraphics[width=0.48\textwidth]{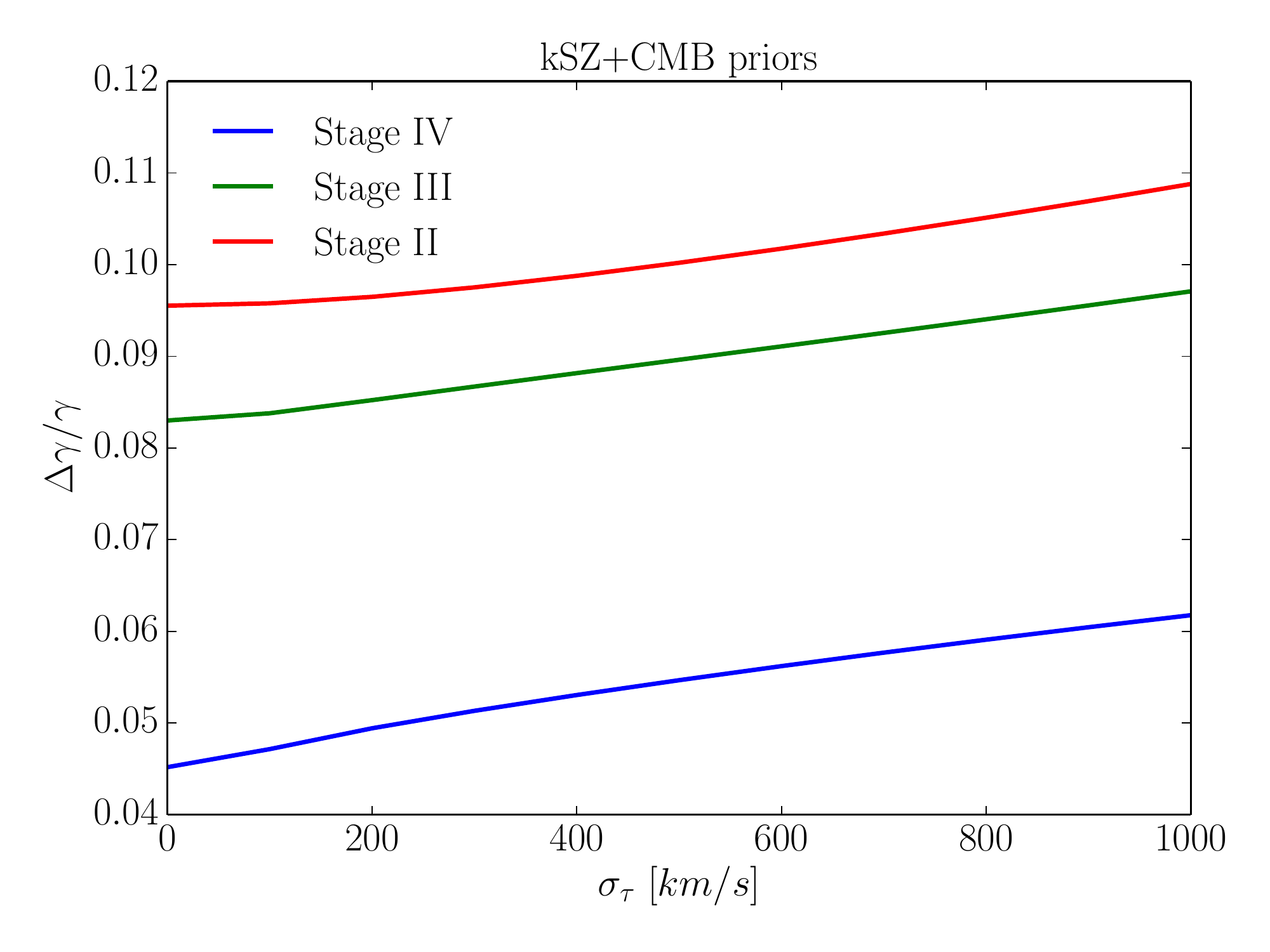}
}
\\
{\includegraphics[width=0.48\textwidth]{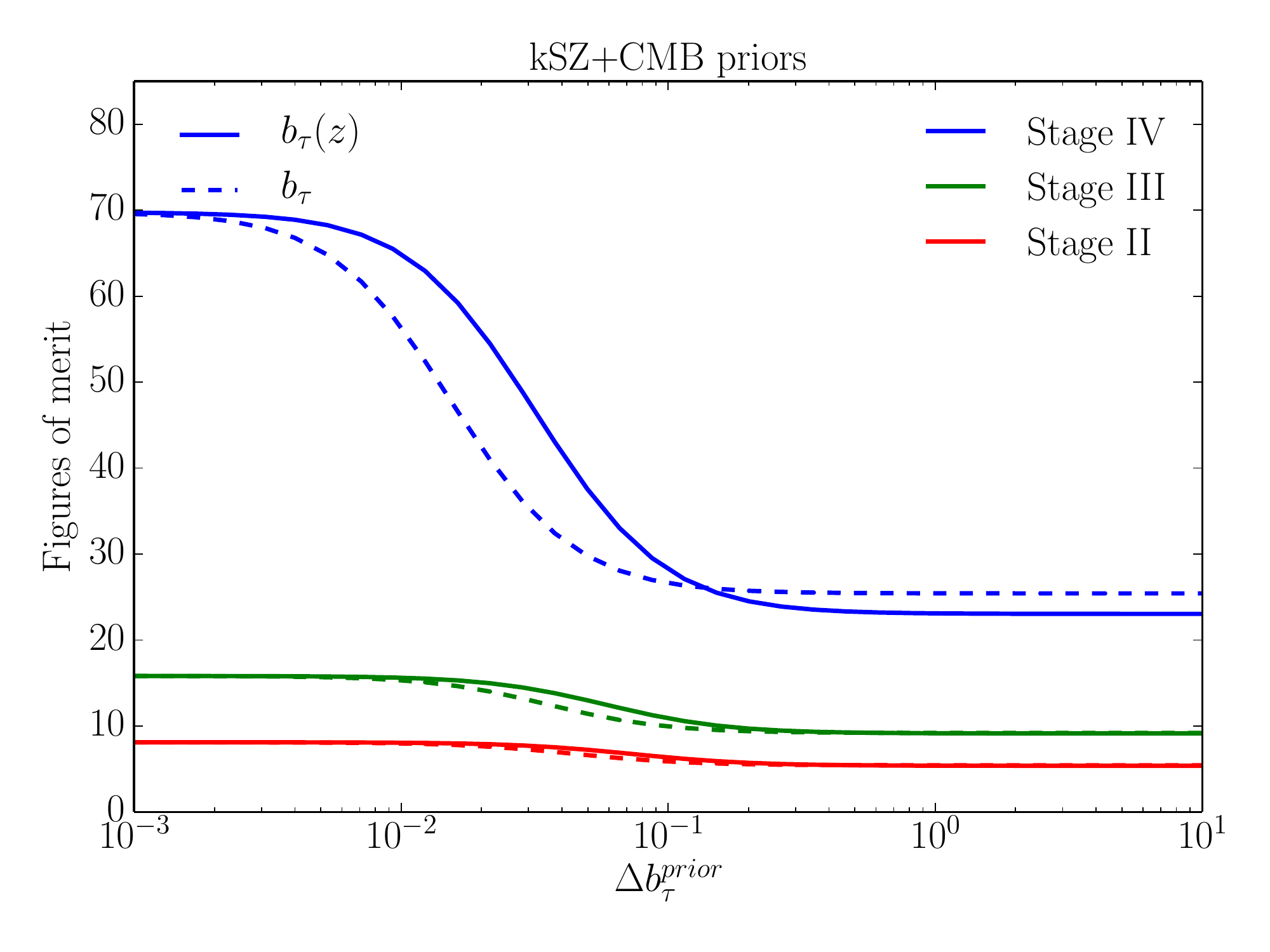} 
\includegraphics[width=0.48\textwidth]{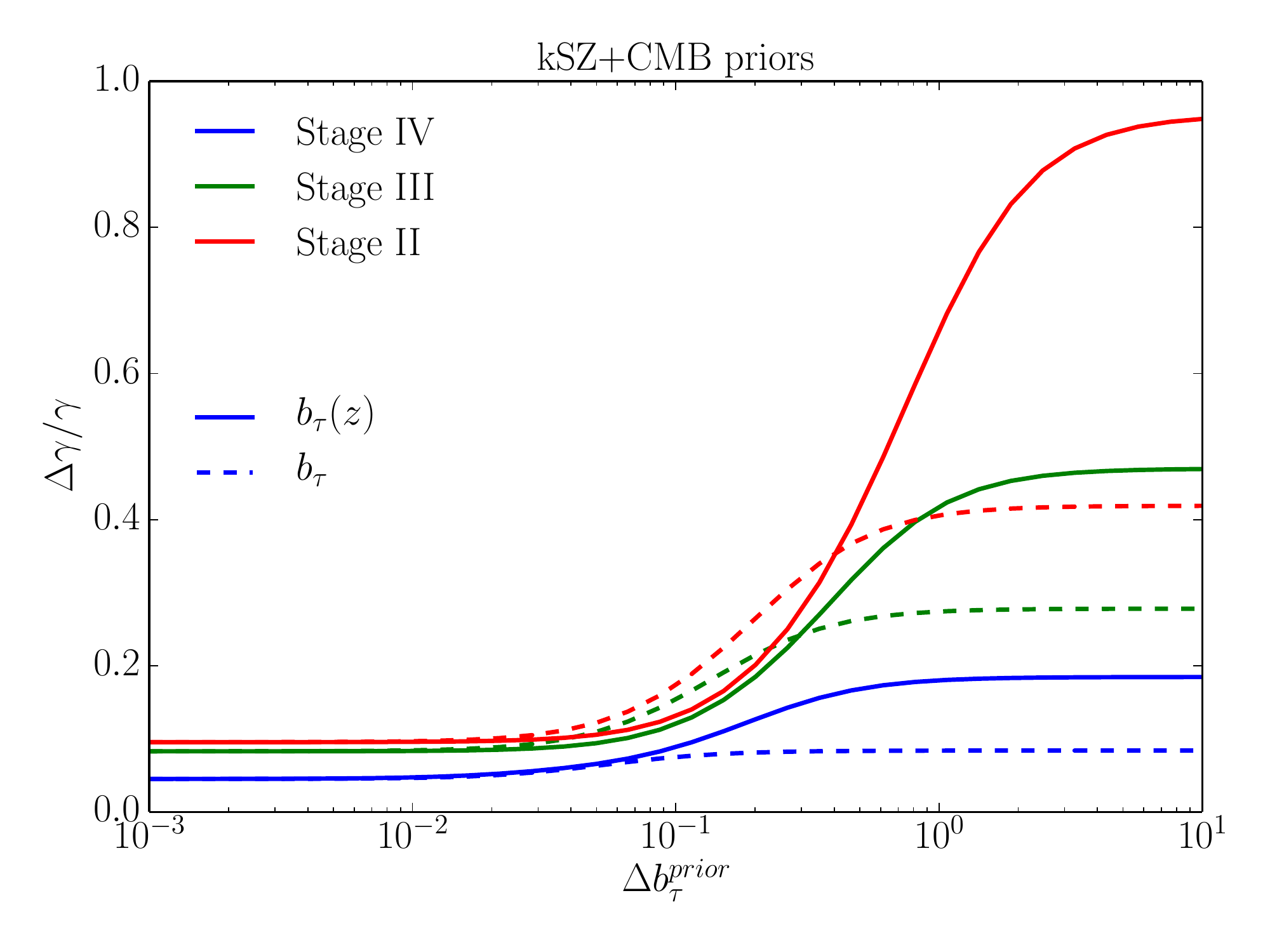}
}
\caption{The impact of modeling assumptions in the determination of $\tau$ for each cluster on the dark energy FoM [left panels] and fractional constraints on the growth factor [right panels] for Stage II (red), Stage III (green) and Stage IV (blue) 
 surveys. The upper panels show the effect of increasing a statistical dispersion in the $\tau$ measurement, $\sigma_\tau$, in the pairwise velocity covariance. The upper left shows the FoM assuming the growth rate is determined by GR (solid lines) and marginalizing over a freely varying $\gamma$ (`MG', dashed lines), which corresponds to the $\Delta \gamma/\gamma$ constraints in the upper right panel. The lower panels show the effect of a prior on a systematic offset in the $\tau$ value, parameterized by a multiplicative bias in each redshift bin (solid lines), $b_{\tau}(z)$, and a redshift independent multiplicative bias (dashed lines), $b_{\tau}$. A detailed discussion
 of the relative sensitivities is provided in the text.}
\label{fig:btau}
\ec
\end{figure*}
%=================================================================
%%%%%%%%%%%%%%%%%%%%%%%%%%%%%%%%%%%%%%%%%%%%
\subsection{Dependence on the non-linearity cut-off}
\label{sec:rcut}
%%%%%%%%%%%%%%%%%%%%%%%%%%%%%%%%%%%%%%%%%%%%
As shown in Figure \ref{fig:CovBin}, the inverse covariance rises 
at lower cluster separations so that the inclusion of cluster pairs at small separation can have a potentially significant effect on improving the dark energy constraints.  Simulation show a deviation from the predicted theoretical mean pairwise velocity, however, starting at separations of $r<45 \  \mathrm{Mpc/h}$ \cite{Bhattacharya:2007sk} so that the  non-linear corrections to the cluster motion needs to be considered. Equation (\ref{eq:vij}) has two major deficiencies: It relies on linear theory to model the underlying dark matter distribution \cite{Juszkiewicz:1998xf, Sheth:2000ff}  and it assumes a linear, scale-independent bias \cite{ Sheth:2000ff}. The former leads to a discrepancy of the dark matter pairwise velocity with linear theory at non-linear scales around $r\leq10 \ \mathrm{Mpc/h}$, the latter introduces deviations at even larger scales.
 It is worthwhile, therefore, to assess how accurately the mean pairwise velocity of clusters can be modeled in the transition to the non-linear regime and how the cosmological constraints depend upon the assumed limiting minimum mass.

In Figure \ref{fig:sspecs} we highlight the sensitivity of the figures of merit and uncertainty in the modified gravity parameter $\gamma$ to the assumptions about the smallest cluster separations to be included in the analysis, parametrized here by $r_{\mathrm{min}}$. For a 
Stage IV like survey including all scales up to $r=5 \ \mathrm{Mpc/h}$ more than doubles the FoM compared to an analysis with separations above $50 \ \mathrm{Mpc/h}$ excluded and halves the uncertainty on $\gamma$. 

In this work we chose a moderate approach cutting off our analysis in the mildly non-linear regime using a minimum separation of $r_{\mathrm{min}}=20$~Mpc$/h$. On-going work on using an perturbative approach to model non-linearities \cite{Okumura:2013zva} and improved N-body simulations suggests that the formalism will be improved in the near future to fully exploit the mildly non-linear regime.

%%%%%%%%%%%%%%%%%%%%%%%%%%%%%%%%%%%%%%%%%%%%
\subsection{Dependence on the measurement error}
\label{sec:sigv}
%%%%%%%%%%%%%%%%%%%%%%%%%%%%%%%%%%%%%%%%%%%%

\label{sec:measure_error}

Central to utilizing the kSZ for cosmology, is the ability to measure the pairwise momentum accurately, and then in turn extract the pairwise velocity, from the momentum, through being able to determine the cluster optical depths. In this section we investigate in more detail the sensitivity of the constraints to these important effects.

As described in section \ref{sec:survey_spec}, the measurement error of a given cluster is given by the combination in quadrature of the instrument noise and the uncertainty in the  optical depth of the cluster. 
In the fiducial analysis we include an uncertainty in the measurement of $\tau$ based on the intrinsic dispersion in the optical depth observed in cluster simulations, averaged over all masses. While this doesn't include the measurement error in estimating the optical depth, it also does not include additional information in the mass dependence of the optical depth that could reduce the intrinsic dispersion estimator through the creation of a fitting function. Possible ways to estimate $\tau$ beyond the scope of this paper include combining thermal SZ and X-ray observations to break the electron temperature-optical depth degeneracy that will partially affect even multi-frequency arcminute resolution observations \cite{Sehgal:2005cy}. This technique relies on theoretical assumptions and modeling to connect the electron temperature to the X-ray temperature, that need more detailed testing against simulations.
A polarization sensitive stage IV CMB survey may be able to measure $\tau$ by stacking clusters to extract the polarization signal introduced by the scattering, which depends directly on the optical depth (see e.g. \cite{Sazonov:1999zp}).

To understand the impact of greater uncertainty in the determination of $\tau$ on the cosmological constraints, we consider two potential forms of uncertainties, shown in Figure \ref{fig:btau}.
The first is the effect of increased statistical dispersion, $\sigma_\tau$ in the optical depths of the cluster sample and  
the second is a systematic offset in the $\tau$. For the latter, we introduce a nuisance parameter, $b_\tau(z)$, in each redshift bin that scales the amplitude of the mean pairwise velocity, $\hat{\Vij}(z)=b_\tau(z) \Vij(z)$, and consider its effect on cosmological constraints when marginalizing over $b_\tau(z)$. Additionally we consider a constant, redshift independent nuisance parameter, $b_\tau$, that scales the amplitude across all clusters. For clarity, when studying the impact of $b_{\tau}$ we remove the $\sigma_\tau^2$ contribution to the covariance and purely parameterize the uncertainty in $\tau$ through a prior on $b_{\tau}$.
 
 Table \ref{tab:kSZ_surveys} shows that for a near-term Stage II survey the noise will be dominated by the instrument accuracy, for a more sensitive Stage III  both components become comparable, and for a Stage IV survey the velocity accuracy may be limited by the accuracy of $\tau$. This is reflected in the top panels of  Figure \ref{fig:btau}  in which varying the amplitude of $\sigma_\tau$ between 0 and 1000 km/s only minimally changes the constraints on the dark energy FoM and the constraints on $\gamma$ for Stage II and III.

For Stage II and Stage III surveys, conclusions for the  effect of $b_\tau$ on $w_0$ and $w_a$ are similar to those for $\sigma_\tau$. The constraints on these dark energy parameters are principally determined by the Planck-like prior, independent of the kSZ constraints, and uncorrelated with $b_\tau$. For the Stage IV survey the kSZ constraints provide additional constraints  on the equation of state, increasing their correlation with $b_\tau$, and the prior has a more pronounced effect on improving the FoM once below $\Delta b_\tau\lesssim 10^{-1}$. Equivalently Stage II and III are not affected by the assumptions on the $\tau$ bias model; marginalizing over the amplitude in each redshift bin yields similar results to introducing a constant bias factor across all redshifts. For Stage IV slightly larger FoM are achieved for a redshift independent $b_\tau$ model without imposing any prior.

For the growth parameter, which is predominantly constrained by the kSZ data, the model assumptions on the $\tau$ nuisance parameter are more important. Marginalizing over the amplitude in each redshift bin, $b_\tau(z)$, without imposing any prior doubles the uncertainty in $\gamma$ compared to a constant, redshift independent nuisance parameter $b_\tau$. The difference between this behavior and the FoM constraints (shown in Figure \ref{fig:btau} lower panels) indicates that the redshift dependence of the FoM versus $\gamma$ helps to break the degeneracy between them. A prior on the bias $\Delta b_\tau\lesssim 10^{-1}$ leads to a factor of $5$ to $10$ improvement in the parameter constraints for the redshift dependent $b_\tau(z)$ model and a factor of $3$ to $4$ improvement for a constant $b_\tau$. For the Stage IV survey we find that the multiple-redshift bins and improved covariance reduce the degeneracy between the $\tau$ bias parameter and $\gamma$, so that the systematic bias and the growth parameter can be constrained simultaneously by the data, and the prior has less effect.

Beyond uncertainties in $\tau$, the measurement uncertainty also depends on the peculiar velocity of the cluster, see 
(\ref{eq:sigv}). Even though the peculiar velocities of clusters are in principle distributed over a range of velocities, here for simplicity we assume a rms velocity of $v=300$ km/s that corresponds to the peak velocity of the distribution found in simulations \cite{Sheth:2000ii} for all clusters to calculate the total measurement error. Fortunately the peak velocity does not strongly depend on the mass of the cluster  \cite{Sheth:2000ii}.  
While future observations will reduce the velocity measurement uncertainty, there is an irreducible error of around $\sigma_v=(50-100)$ km/s on the cluster peculiar motion due to internal motion within the cluster \cite{Nagai:2002nw,Holder:2002wc} that will ultimately limit the CMB observations.

 %=================================================================
\begin{figure}[!t]
\bc
{\includegraphics[width=0.48\textwidth]{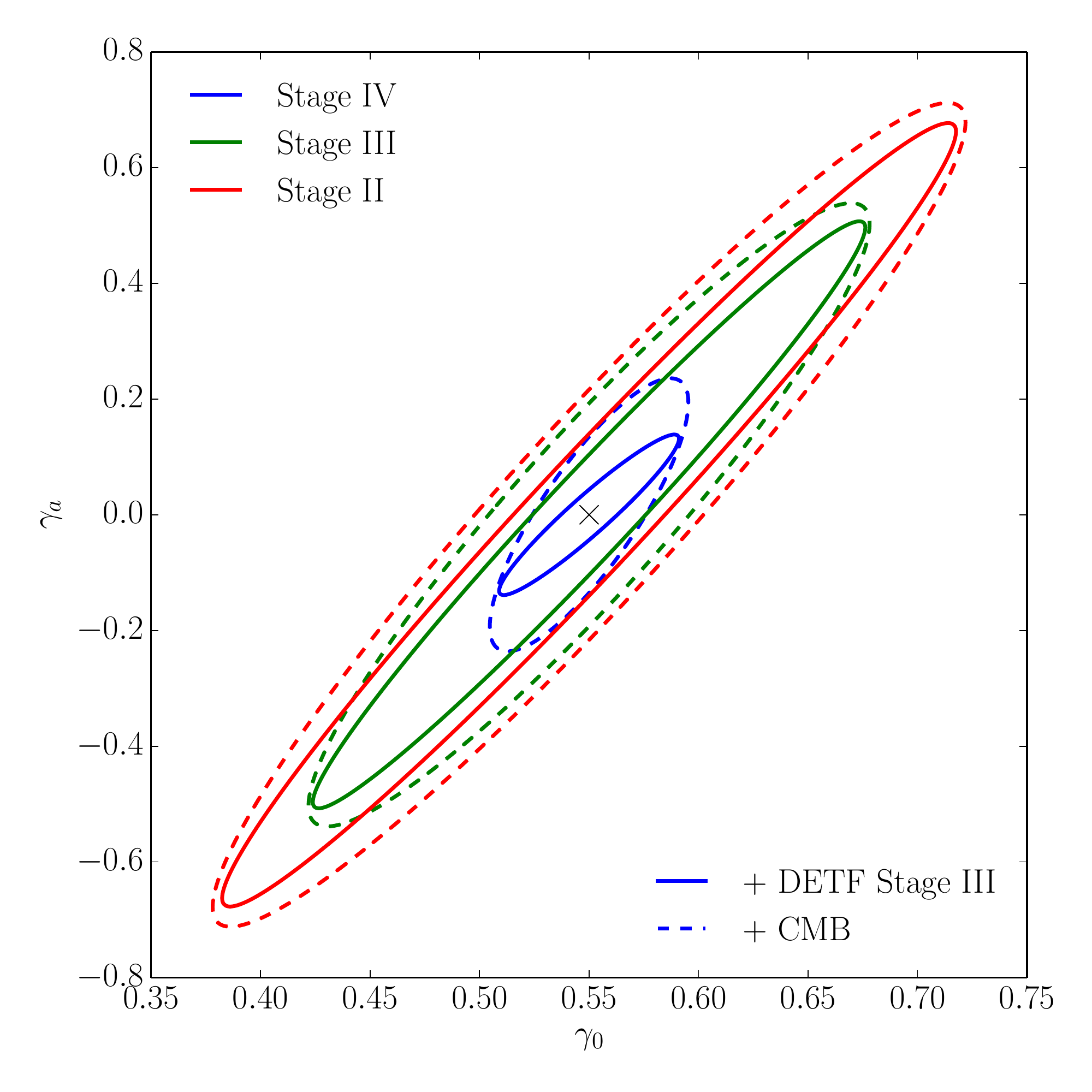}
}
\caption{Marginalized constraints on the $\gamma_0 - \gamma_a$ parameter space showing the 68\% confidence contours for the pairwise velocity constraints in combination with a  Planck-like  (dashed) or DETF Stage III (solid) prior on all parameters except $\gamma_0$ and $\gamma_a$. 
The fiducial model assumes GR  with $\gamma_0=0.55$ and $\gamma_a=0$.}
\label{fig:2D_g0ga}
\ec
\end{figure}
%=================================================================

\begin{table}[!t]
\begin{center}
\begin{tabular}{|l|r|r|r|}
        \cline{2-4}
\multicolumn{1}{c}{}& \multicolumn{3}{|c|}{ $\gamma=\gamma_0+\gamma_a(1-a)$ }\\
\cline{2-4}
 \multicolumn{1}{c|}{}                          &   Stage II &   Stage III &   Stage IV \\
\hline

 $\mathrm{FoM}_{\mathrm{MG}}$                &      130   &       151     &     269   \\
 $\sigma(w_0)$             &      0.10  &       0.08  &     0.06 \\ 
 $\sigma(w_a)$             &      0.29 &       0.26 &      0.21 \\
 $\Delta\gamma_0/\gamma_0$ &      0.31 &       0.23 &      0.08 \\
 $\sigma(\gamma_a)$        &      0.68 &       0.51 &      0.14 \\
\hline
\end{tabular}

    \caption{A summary of the dark energy FoM and 1$\sigma$ marginalized constraints on for the dark energy parameters in the $\gamma_0-\gamma_a$ parametrization for Stage II, III and IV scenarios in combination with a DETF  prior on all parameters except $\gamma_0$ and $\gamma_a$.  \label{tab:results_g0ga}}
  \end{center}
\end{table}

%=================================================================
\begin{figure}[!t]
\bc
\vspace{-0.6cm}
{\includegraphics[width=0.51\textwidth]{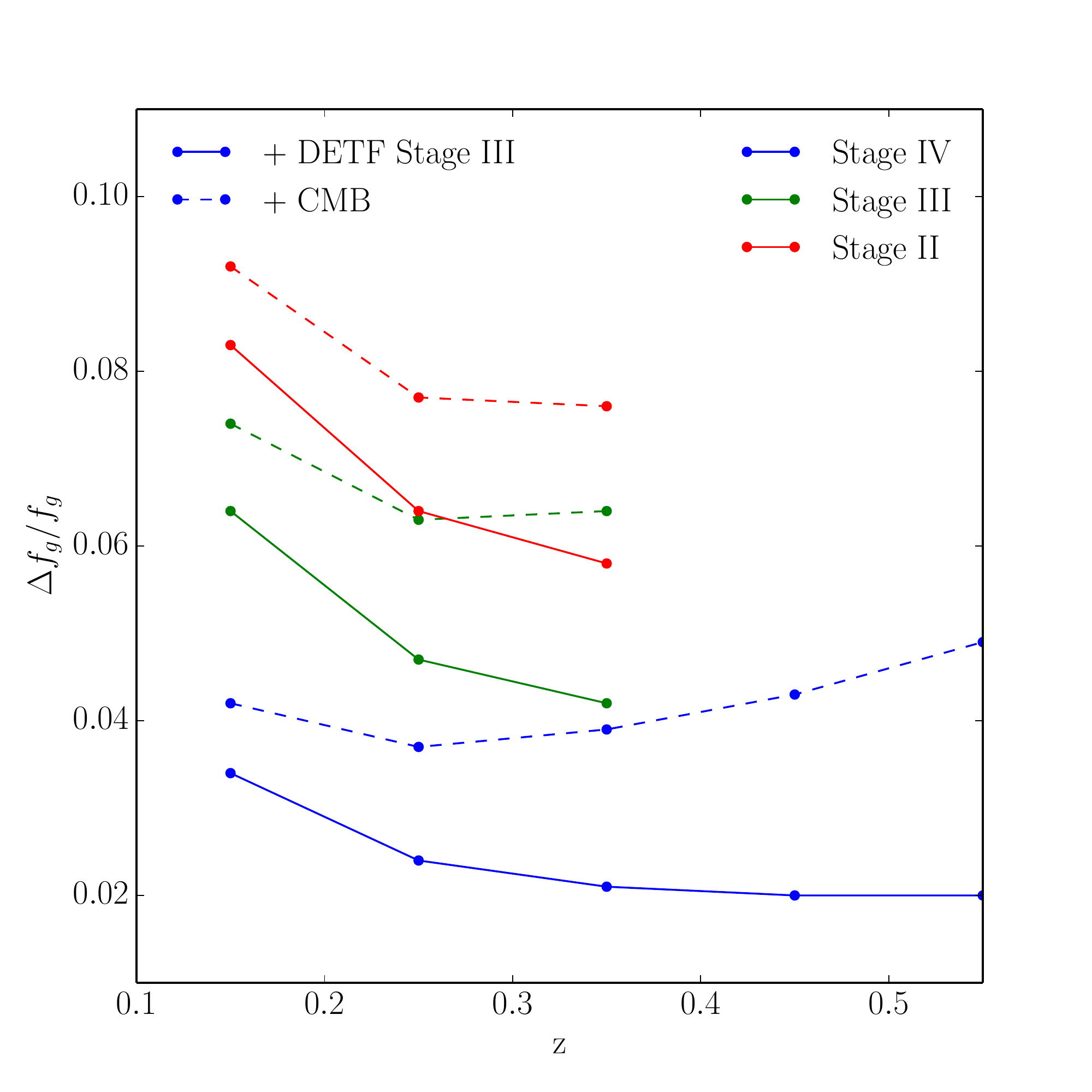} 
}
\caption{Expected fractional $1\sigma$ errors on the growth rate, $f_g$, in each redshift bin for Stage II (red), Stage III (green) and Stage IV (blue) when combined with a Planck-like CMB (dashed) or DETF Stage III (solid) prior on all parameters except $f_g(z)$.} 
\label{fig:fg} 
\ec
\end{figure}
%=================================================================
%%%%%%%%%%%%%%%%%%%%%%%%%%%%%%%%%%%%%%%%%%%%
\subsection{Dependency on the Modified Gravity parametrization}
\label{sec:deparam}
%%%%%%%%%%%%%%%%%%%%%%%%%%%%%%%%%%%%%%%%%%%%
In the previous sections we parametrized modified gravity models using one extra parameter $\gamma$ that is assumed constant across all redshifts. Not all modified gravity models are well represented by such a simple parameterization. Some models are better fit by a more general parametrization that allows for a monotonic redshift dependence in $\gamma$, $\gamma(a)=\gamma_0+(1-a)\gamma_a$ \cite{Wu:2009zy}, equivalent to the dark energy $w_0-w_a$ model. Table \ref{tab:results_g0ga}
summarizes the figures of merit as well as the $1\sigma$ constraints on \{$w_0$,$w_a$,$\gamma_0$,$\gamma_a$\}. Introducing an additional extra parameter  loosens the constraints on the parameters with the advantage of imposing a smaller theoretical prior on modified gravity. Figure \ref{fig:2D_g0ga} shows the $1\sigma$ and $2\sigma$ constraints on $\gamma_0-\gamma_a$ for Stage II, III and IV.

An even more general approach is to directly constrain the growth rate in redshift bins as a `model-independent' way. This approach is particularly applicable for spectroscopic galaxy surveys which can isolate peculiar velocity data, and hence the growth rate, in  precise redshift bins.  In Figure \ref{fig:fg} we present the forecasts for this parametrization. We find, in combination with  the Stage III DETF constraints on the equation of state, an uncertainty in $f_g$ of less than 2\% at $z\sim 0.3-0.6$ for Stage IV surveys. These constraints provide a complementary measurement to the growth rate measurements from redshift space clustering of emission line galaxies (ELGs) and LRGs  with a DESI-like survey. On its own, without the additional DETF stage III data, forecasts for DESI project 1.4-1.6\% errors on the growth factor multiplied by $\sigma_8$ over a comparable redshift range to that considered in our analysis \cite{Levi:2013gra}. DESI, along with Euclid \cite{Amendola:2012ys} and WFIRST \cite{Spergel:2013tha} will also provide complementary spectroscopic constraints on the growth rate at higher redshifts, $1<z<3$.

%%%%%%%%%%%%%%%%%%%%%%%%%%%%%%%%%%%%%%%%%%%%
\section{Conclusions }
\label{sec:conclusions}
%%%%%%%%%%%%%%%%%%%%%%%%%%%%%%%%%%%%%%%%%%%%

Recent analyses have demonstrated that the kSZ can be successfully extracted from sub arcminute resolution CMB maps by cross-correlating them with cluster positions and redshift from spectroscopic large scale structure surveys. In this paper we have considered the potential to apply this technique, in light of planned CMB and LSS surveys with greater sensitivity and larger areas,  to constrain dark energy and modifications to gravity on cosmic scales using the mean pairwise velocity of clusters as an observable.
We have extended the model presented in \cite{Bhattacharya:2007sk} to account for the dependence on the binning in cluster separations, shot noise, and potential 
contributions to the total covariance matrix due to small number densities, that we show are significant, despite being frequently neglected, and provided a detailed derivation of the covariance components.

The projected constraints are intimately related to not only the quality of future data, determined by the instrumental precision, but also to the modeling of the uncertainties in transforming the kSZ observations into velocity estimates that constrain the large scale structure growth history. We investigate a range of  uncertainties, using reasonable assumptions based on simulations and projected survey capabilities. We also study the sensitivity to assumptions by 
varying theoretical priors to understand and estimate the robustness of the results. We included a study of  the effect of survey assumptions on the minimum detectable cluster mass and the minimum cluster separation that could be included, in light of the influence of non-linear effects in the cluster motions/correlations.

The mean pairwise velocity is modeled assuming linear theory for the underlying matter distribution as well as the halo bias. Variations in the equation of state and the growth rate affect the linear growth factor in similar ways, that leads to  degenerate effects on the pairwise velocity amplitude. However, the different redshift dependence of these effects helps to break the degeneracy, and constraints on the expansion history, such as those from Type 1a supernovae, BAO and CMB geometric constraints, break it further, allowing growth information to be extracted from the kSZ.

The cluster sample's minimum mass  has a significant impact on the predicted constraints. A smaller minimum mass leads to an increase of the number of clusters in the catalog (assuming the catalog is nearly complete) and significantly
 reduces the errors on all cosmological parameters. Assuming an optimistic mass cut-off for the upcoming cluster catalogs leads to an improvement on the figures of merit (including CMB priors) from $\mathrm{FoM}_{\mathrm{GR}}=61$ to $\mathrm{FoM}_{\mathrm{GR}}=110$ and a reduction of the 1$\sigma$ uncertainty of the modified gravity parameter $\gamma$ from $5\%$ to $2\%$ for a Stage IV survey compared to 
our fiducial assumption.

In contrast, the uncertainty in  the exact minimum mass had only a mild impact on the dark energy and modified gravity constraints. This was understood in terms of the additional effect of the minimum mass on the shape, as a function of cluster pair separation, as well as amplitude, of the pairwise velocity. Marginalizing over the minimum mass, while imposing a $15\%$ prior in our analysis, in a scenario in which the covariance remains unchanged, reduces the FoM for a Stage IV survey from $\mathrm{FoM}_{\mathrm{GR}}=61$ to $\mathrm{FoM}_{\mathrm{GR}}=43$ and marginally loosens the constrains on $\gamma$ since the mean pairwise velocity is only weakly dependent on the assumed mass cut-off. In comparison to the abundance of clusters as a cosmological probe, the mean pairwise velocity of clusters appears to be more robust to uncertainties in the mass calibration.

Considering pairwise correlations down to cluster separations of $r=5\mathrm{Mpc/h}$ doubles the FoM compared to an analysis that excludes all scales below $r=50\mathrm{Mpc/h}$. While extending the analysis to smaller separations could significantly improve the constraints, including scales in the non-linear regime without accurate modeling could also potentially bias the constraints and introduce more systematic uncertainties.

Improved constraints on $\tau$ in clusters are critical for accurate extraction of cluster streaming velocities from kSZ measurements.
We studied the impact of uncertainties in the $\tau$ measurement by considering constraints as we varied the level of statistical uncertainty in individual cluster $\tau$ measurements and, separately, the effect of a systematic offset in the $\tau$ determinations. The later was parameterized by a multiplicative bias parameter in each redshift bin, $b_\tau(z)$, as well as a constant, redshift independent bias, $b_\tau$.  We found that the effect of $\sigma_\tau$ on the dark energy FoM was minimal reflecting that the principal constraints come from the external CMB or DETF prior.  For Stage IV, in particular, the dispersion in $\tau$ does have a notable impact on the growth factor constraints as the instrument contribution to the measurement error and shot noise contributions have decreased.
For the systematic offset in $\tau$, we found that the prior on $b_\tau$ or $b_\tau(z)$ had the biggest impact for Stage II and Stage III surveys for which significant degeneracies exist between the $\tau$ bias and $\gamma$. Though, a $\sim$10\% prior on the amplitude of $\tau$ enables these surveys to provide competitive constraints. For a Stage IV survey and a $b_\tau$ bias model, the redshift bins and reduced covariance allowed both $b_\tau$ and $\gamma$
to be extracted from the data without the need for a prior on $\tau$.

In addition to a minimal model to modify gravity, in which a modification to the growth rate is parameterized by a single parameter, $\gamma$, we also predict constraints for more general modified gravity parametrizations. We use a  $\gamma$ parametrization that monotonically varies with the scale factor, and a model independent approach of measuring  the growth rate as a function of redshift, $f_g(z)$, directly. We forecast $\sim 6-8\%$ 1$\sigma$ errors on $f_g$ for Stage II, $4\%-6 \%$ for Stage III and  $\sim$2\% constraints for Stage IV when combined with a Stage III DETF constraints on the expansion history.

Potential improvements in the covariance could include taking advantage of multi-frequency information available in upcoming surveys (e.g. \cite{Calabrese:2014, Benson:2014}) to improve the kSZ signal extraction and reduce the measurement error. Larger LRG catalogs could also be used, 
 such as in the first kSZ detection; however,  this increases uncertainty in the minimum mass of the cluster sample. 
 Similarly, with the improvements in cluster photometric redshift uncertainties that are coming from improved algorithms and spectroscopic training sets, it may be feasible to use photometric surveys, without spectroscopic follow up, to significantly enlarge the cluster sample. This will degrade the redshift accuracy, and therefore the measurements of the cluster separation, particularly on small scales; however, the larger sample size will help compensate and might even improve the constraining power.
 
Measurements of the kSZ effect provide complementary constraints on the growth of structure to weak lensing and redshift space distortion measurements by providing measurements on larger physical scales and using a highly complementary, and more massive, tracer of the cosmological gravitational field, that is not dependent upon a characterization of galaxy bias. Having a variety of cosmological probes of dark energy and modified gravity with different systematics  is going to be vital for reducing systematic effects and biases in parameter estimation and determining the properties of dark energy and gravity in a variety of epochs and regimes.

%%%%%%%%%%%%%%%%%%%%%%%%%%%%%%%%%%%%%%%%%%%%
\section*{Acknowledgments}
%%%%%%%%%%%%%%%%%%%%%%%%%%%%%%%%%%%%%%%%%%%%
The authors would like to thank Nicholas Battaglia, Joanna Dunkley, Kira Hicks, Arthur Kosowsky, Thomas Loredo, Eduardo Rozo and David Spergel for useful inputs and discussions on pairwise statistics, survey capabilities and astrophysical uncertainties, and comments on the manuscript. 

The work of EMM and RB is supported by NASA ATP grants NNX11AI95G and NNX14AH53G, NASA ROSES grant 12-EUCLID12- 0004, NSF CAREER grant 0844825 and DoE grant DE-SC0011838.

\onecolumngrid
\appendix
\section{Derivation of the  mean pairwise velocity covariance}\label{app:cov}

In this section we provide a detailed derivation of the Gaussian and non-Gaussian contributions to the covariance matrix given in equation (\ref{eq:shot}) in the main text.  The covariance matrix specified in (\ref{eq:covdef}) in terms of the volume average of the estimator, $\hat{\Vij}$ of the pairwise velocity $\Vij$,
\bea
C_{\Vij}(r,r')=\langle \hat{\Vij}(r)\hat{\Vij}(r')\rangle-\langle \hat{\Vij}(r) \rangle \langle \hat{\Vij}(r') \rangle.
\eea
Let's first consider a covariance between the pairwise cluster velocities of two cluster pairs, each with respective separations $r$ and $r'$, we will then incorporate the effect of including finite bin sizes in the cluster separations.
Using the expression for the mean pairwise cluster velocity, given in (\ref{eq:vij}), the covariance of $\Vij$  can be written as
\bea
C_{\Vij}(r,a,r',a')&=&\frac{1}{1+\xi_{h}(r,a)} \frac{2}{3} r H(a) a f_g(a) \frac{1}{1+\xi_{h}(r',a')}\frac{2}{3} r' H(a') a' f_g(a') \\
& \times& \left[ \langle \hat{\bxi}_{h}(r)\hat{\bxi}_{h}(r')\rangle-\langle \hat{\bxi}_{h}(r) \rangle \langle \hat{\bxi}_{h}(r') \rangle \right].
\eea

For simplicity in the following derivation, we drop the subscript ``h" (denoting halo) from the correlation function, $\xi_{h}$ and mass average correlation function, $\bar\xi_{h}$, denoting them respectively by $\xi$ and $\bxi$. Similarly we use $P(k,a)$ to denote the halo linear dark matter power spectrum, given in full by $P_{lin}^{dm}(k,a) b^{(2)}_{h}(k)$, and the cluster number density $n_{cl}$ is denoted $n$.

We define an estimator of the volume averaged correlation function $\bxi$ equivalently to the estimator of the correlation function $\xi$ as
 \bea
\hat{\bar{\xi}}(\vec{r})&=&\frac{1}{V(\vec{r})} \int_0^{\vec{r}} d^3r_1  \frac{1}{V(\vec{r}_1)}\int d^3x W(\vec{x})\int d^3x'W(\vec{x}')\delta(\vec{x})\delta(\vec{x}') \delta_D^{(3)}(\vec{x}-\vec{x}'-\vec{r}_1) \\
&=&\frac{1}{V(\vec{r})} \int_0^{\vec{r}_1} d^3r_1 \int\frac{d^3k}{(2\pi)^3}\int\frac{d^3k_1}{(2\pi)^3}\delta_{\vec{k}}\delta^*_{\vec{k_1}}e^{i\vec{k}_1 \vec{r}} h(\vec{k}-\vec{k}_1,\vec{r}_1)
\eea
where
\bea
h(\vec{k},\vec{r})=\frac{1}{V(\vec{r})}\int d^3x e^{i\vec{k}\vec{r}} W(\vec{x})W(\vec{x}+\vec{r}).
\eea
The covariance matrix at a given redshift (dropping the subscript a) becomes
\bea
C_{\bar{\xi}}(\vec{r},\vec{r}')&=&\frac{1}{V(\vec{r})} \int_0^{\vec{r}} d^3r_1 \frac{1}{V(\vec{r}')} \int_0^{\vec{r}'} d^3r'_1 \int\frac{d^3k}{(2\pi)^3}\int\frac{d^3k_1}{(2\pi)^3}e^{i\vec{k}_1 \vec{r}_1} h(\vec{k}-\vec{k}_1,\vec{r}_1)  \nonumber \\
&\times&\int\frac{d^3k'}{(2\pi)^3}\int\frac{d^3k'_1}{(2\pi)^3}e^{i\vec{k'}_1 \vec{r'}_1} h(\vec{k'}-\vec{k'}_1,\vec{r}'_1) \nonumber \\
&\times& \left[ \langle \delta_{\vec{k}} \delta^*_{\vec{k}_1}\delta_{\vec{k'}} \delta^*_{\vec{k}'_1} \rangle -  \langle  \delta_{\vec{k}} \delta^*_{\vec{k}_1} \rangle \langle\delta_{\vec{k'}} \delta^*_{\vec{k}'_1} \rangle \right].
\eea
The expectation value of the four $\delta's$ including noise is
\bea
 &&\left[ \langle \delta_{\vec{k}} \delta^*_{\vec{k}_1}\delta_{\vec{k'}} \delta^*_{\vec{k}'_1} \rangle -  \langle  \delta_{\vec{k}} \delta^*_{\vec{k}_1} \rangle \langle\delta_{\vec{k'}} \delta^*_{\vec{k}'_1} \rangle \right] \\
 &=& (2\pi)^3 \delta_D^{(3)}(\vec{k}+\vec{k}') \left( P(\vec{k})+\frac{1}{n} \right)(2\pi)^3 \delta_D^{(3)}(\vec{k}_1+\vec{k}_1') \left( P(\vec{k})_1+\frac{1}{n} \right)\\
 &+& (2\pi)^3 \delta_D^{(3)}(\vec{k}-\vec{k}_1') \left( P(\vec{k})+\frac{1}{n} \right)(2\pi)^3 \delta_D^{(3)}(\vec{k}_1-\vec{k}') \left( P(\vec{k})_1+\frac{1}{n} \right)\\
 &+&(2\pi)^3\delta_D^{(3)}(\vec{k}-\vec{k}_1+\vec{k}'-\vec{k}_1') T^{full}_4(\vec{k},\vec{k}_1,\vec{k}',\vec{k}'_1).
\eea
Evaluating the first two terms of the above equation leads to the Gaussian contribution of the covariance matrix
\bea
C_{\xi}(\vec{r},\vec{r}')&=&\frac{1}{V(\vec{r})} \int_0^{\vec{r}} d^3r_1 \frac{1}{V(\vec{r}')} \int_0^{\vec{r}'} d^3r'_1 \int\frac{d^3k}{(2\pi)^3}\int\frac{d^3k_1}{(2\pi)^3}\left( P(\vec{k})+\frac{1}{n} \right)\left( P(\vec{k}_1)+\frac{1}{n} \right)
\\
&\times&h(\vec{k}-\vec{k}_1,\vec{r})h^*(\vec{k}'-\vec{k}'_1,\vec{r}')\left( e^{i\vec{k}_1\vec{r}+i\vec{k}\vec{r}'}+e^{i\vec{k}_1\vec{r}-i\vec{k}_1\vec{r}'} \right) \nonumber. \\
\eea
Using the approximation
\bea
\int \frac{d^3k}{(2\pi)^3} h(\vec{k},\vec{r})h^*(\vec{k},\vec{r}')=\frac{1}{V_s(a)}\delta_{aa'}
\eea
the Gaussian terms become
\bea
C_{\bar{\xi}}(\vec{r},\vec{r}')
&=&\frac{1}{V(\vec{r})} \int_0^{\vec{r}} d^3r_1 \frac{1}{V(\vec{r}')} \int_0^{\vec{r}'}  d^3r'_1 \frac{1}{V_s}\int \frac{d^3k}{(2\pi)^3} \left( P(\vec{k})^2+\frac{2 P(\vec{k})}{n}+\frac{1}{n^2} \right) \left[ e^{i\vec{k}(\vec{r}_1+\vec{r'}_1)}+e^{i\vec{k}(\vec{r}_1-\vec{r}_1')} \right] \\
&=& \frac{1}{\pi^2 V_s(a)}\frac{1}{V(r)V(r')}  \int k^2 dk  \int_0^r 4 \pi r_1^2  dr_1 \int_0^{r'} 4\pi  r_1^{\prime 2} dr'_1  \left( \frac{\sin(kr_1)}{kr_1}\right) \left( \frac{\sin(kr'_1)}{kr'_1}\right) P(k)^2 \\
&=& \frac{9}{\pi^2 r r'  V_s(a)}\int dk \left( P(k)^2+\frac{2 P(k)}{n}+\frac{1}{n^2} \right) j_1(kr) j_1(k r').
\eea

The remaining non-Gaussian terms come from the trispectrum \cite{Matarrese:1997sk}
\bea
T^{full}_4(\vec{k},\vec{k}_1,\vec{k}',\vec{k}'_1)&=& \frac{1}{n^2}\left[ P(\vec{k}-\vec{k}_1+\vec{k}') + P(\vec{k}+\vec{k}'-\vec{k}'_1) + P(\vec{k} -\vec{k}'_1-\vec{k}_1) + P(\vec{k}'-\vec{k}_1-\vec{k}_1') \right] \nonumber \\
&+& \frac{1}{n^2}\left[ P(\vec{k}-\vec{k}_1)+P(\vec{k}+\vec{k}')+P(\vec{k}-\vec{k}'_1)\right] + \frac{1}{n^3} \label{eq:trispectrum}
\eea
dropping all the terms proportional to the bispectrum, and four point functions, assuming a Gaussian density distribution.

Evaluating the first term of (\ref{eq:trispectrum})  leads to a non-zero contribution only for a separation with $r=0$, proportional to $ \left[ 2 \xi(\vec{r}) \delta_D^{(3)}(\vec{r}') + 2 \xi(\vec{r}') \delta_D^{(3)}(\vec{r}) \right] $ and is therefore not relevant for this work. Similarly, the last term leads to
\bea
&&\int\frac{d^3k}{(2\pi)^3}\int\frac{d^3k_1}{(2\pi)^3}e^{i\vec{k}_1 \vec{r}} h(\vec{k}-\vec{k}_1,\vec{r})
\int\frac{d^3k'}{(2\pi)^3}\int\frac{d^3k'_1}{(2\pi)^3}e^{i\vec{k'}_1 \vec{r'}} h(\vec{k'}-\vec{k'}_1,\vec{r'}) \frac{1}{n^3}  \nonumber \\
&\times& (2\pi)^3\delta_D^{(3)}(\vec{k}-\vec{k}_1+\vec{k}'-\vec{k}_1') \nonumber \\
&=&\frac{1}{n^3 V_s}\delta_D^{(3)}(\vec{r}) \delta_D^{(3)}(\vec{r}').
\eea
The only non-zero term, proportional to $1/n^2$, gives rise to a non-Gaussian contribution to the covariance, we will denote as 'Poisson' shot noise term, and can be evaluated using
\bea
&&\int\frac{d^3k}{(2\pi)^3}\int\frac{d^3k_1}{(2\pi)^3}e^{i\vec{k}_1 \vec{r}} h(\vec{k}-\vec{k}_1,\vec{r})
\int\frac{d^3k'}{(2\pi)^3}\int\frac{d^3k'_1}{(2\pi)^3}e^{i\vec{k'}_1 \vec{r'}} h(\vec{k'}-\vec{k'}_1,\vec{r'})
\times (2\pi)^3\delta_D^{(3)}(\vec{k}-\vec{k}_1+\vec{k}'-\vec{k}_1') \nonumber \\
&\times& \left(\frac{1}{n^2}\left[ P(\vec{k}-\vec{k}_1)+P(\vec{k}+\vec{k}')+P(\vec{k}-\vec{k}'_1)\right] \right) \\
&=&\int\frac{d^3k}{(2\pi)^3}\int\frac{d^3k_1}{(2\pi)^3}e^{i\vec{k}_1 \vec{r}} h(\vec{k}-\vec{k}_1,\vec{r})
\int\frac{d^3k'}{(2\pi)^3}e^{i(\vec{k}-\vec{k}_1+\vec{k}') \vec{r'}} h(\vec{k}_1-\vec{k},\vec{r'}) \nonumber \\
&\times& \left(\frac{1}{n^2}\left[ P(\vec{k}-\vec{k}_1)+P(\vec{k}+\vec{k}')+P(\vec{k}_1-\vec{k}')\right] \right) \\
&=&\frac{1}{n^2 V_s}\int\frac{d^3k}{(2\pi)^3}\int\frac{d^3k'}{(2\pi)^3}e^{i \vec{k}\vec{r}} e^{i \vec{k}' \vec{r}'}\left[ P(\vec{k}+\vec{k}')+P(\vec{k}-\vec{k}')\right] \\
&=&\frac{1}{n^2 V_s}\int\frac{d^3k}{(2\pi)^3}\int\frac{d^3k'}{(2\pi)^3}e^{i \vec{k}'\vec{r}} e^{i (\vec{k}-\vec{k}') \vec{r}'} P(\vec{k}) +\frac{1}{n^2 V_s}\int\frac{d^3k}{(2\pi)^3}\int\frac{d^3k'}{(2\pi)^3}e^{i (\vec{k}+\vec{k}')\vec{r}} e^{i \vec{k}' \vec{r}'} P(k) \\
&=&\frac{1}{n^2 V_s}\int\frac{d^3k}{(2\pi)^3}\int\frac{d^3k'}{(2\pi)^3}e^{i \vec{k}'(\vec{r}-\vec{r}')} e^{i \vec{k} \vec{r}'} P(\vec{k}) +\frac{1}{n^2 V_s}\int\frac{d^3k}{(2\pi)^3}\int\frac{d^3k'}{(2\pi)^3}e^{i \vec{k}\vec{r}} e^{i \vec{k}' (\vec{r}+\vec{r}')} P(k) \\
&=&\frac{1}{n^2 V_s}(\delta_D^{(3)}(\vec{r}-\vec{r}') \xi(\vec{r}) + \delta_D^{(3)}(\vec{r}+\vec{r}') \xi(\vec{r}) )
\eea
dropping the term proportional to $P(0)$.
Using spherical symmetry
\bea
\delta_D(\vec{r}-\vec{r}') \xi(\vec{r}') + \delta_D(\vec{r}'-\vec{r}) \xi(\vec{r})=  \frac{\delta_D(r-r')}{4\pi r^2} \xi(r') + \frac{\delta_D(r'-r)}{4 \pi r'^2} \xi(r)
\eea
and volume averaging over $r$ and $r'$ leads to
\bea
\frac{1}{V(r)}\frac{1}{V(r')} \int_0^{r'} \int_0^r \left( \frac{\delta_D(\tr-\tr')}{4\pi \tr^2} \xi(\tr') + \frac{\delta_D(\tr'-\tr)}{4 \pi \tr'^2} \xi(\tr) \right) 4\pi \tr^2 4 \pi \tr'^2 d\tr d\tr' \\
=   \left\{
  \begin{array}{l l}
    \frac{2}{V(r')} \bar{\xi}(r) & \quad \text{if $r'>r$ }\\
    \frac{2}{V(r)} \bar{\xi}(r')  & \quad \text{if $r'<r$} \\
        \frac{1}{V(r')} \bar{\xi}(r)  +  \frac{1}{V(r)} \bar{\xi}(r') & \quad \text{if $r'=r$}
  \end{array} \right. \\
\eea
where we have used the integral expression for the Dirac delta function
\bea
\int_0^r \delta_D(\vec{\tr}-\vec{\tr}') d^3r&=&\int_0^r \int_{-1}^1 2 \pi \tr^2 d\tr d\mu \int \frac{dk}{(2\pi)^3} \int_{-1}^1 d\mu' 2 \pi k^2 e^{ik\mu \tr-ik\mu'r'} \\
&=&\frac{2}{\pi} \int dk k j_1(kr)j_0(kr') r^2
\eea
to calculate the integrals as
\bea
\int_0^r \int_0^{r'} \delta_D(\vec{\tr}-\vec{\tr}') \xi(\vec{\tr}') d^3\tr d^3\tr'&=&\int_0^{r'} \frac{2}{\pi} \int dk k j_1(kr)j_0(k\tr') r^2 \frac{1}{2 \pi^2}\int dk' k'^2 j_0(k'\tr') P(k') 4 \pi \tr'^2 d\tr' \nonumber \\
&=& \frac{2}{\pi} \int dk' P(k')  \left\{
 \begin{array}{l l}
    j_1(rk') k' r^2 & \quad \text{if $r'>r$ }\\
    j_1(r'k') k' r'^2 & \quad \text{if $r'<r$} \\
    \frac{1}{2} j_1(rk') k' r^2 + \frac{1}{2}  j_1(r'k') k' r'^2 & \quad \text{if $r'=r$}
  \end{array} \right. \\
\eea

The total covariance for the pairwise velocity correlation of cluster pairs of exact separation $r$ and $r'$ can therefore be written as
\bea
&&C_{\Vij}(r,a,r',a')= C_{\Vij}^{Gaus.}(r,a,r',a') + C_{\Vij}^{Poiss.} (r,a,r',a')\eea
with
\bea
 C_{\Vij}^{Gaus.}(r,a,r',a') &=& \frac{4}{\pi^2 V_s(a)} \frac{H(a)a}{1+\xi(r,a)}\frac{H(a')a'}{1+\xi(r',a')}  f_g(a) f_g(a')\ \delta_{aa'} \int dk \left( P(k,a)+\frac{1}{n(a)} \right) \left( P(k,a')+\frac{1}{n(a')} \right) \nonumber \\
 &\times&  j_1(kr) j_1(k r') 
\hspace{0.5cm} 
 \\ \nonumber \\
C_{\Vij}^{Poiss.} (r,a,r',a') &=& \frac{1}{ V_s(a)} \frac{1}{3 \pi}\frac{H(a)a}{1+\xi(r,a)}\frac{H(a')a'}{1+\xi(r',a')}  f_g(a) f_g(a')\ \delta_{aa'}  \left\{
  \begin{array}{l l}
    \frac{r}{n(a)^2 r'^2} \bar{\xi}(r,a) & \quad \text{if $r'\geq r$ }\\
    \frac{2r'}{n(a')^2r^2} \bar{\xi}(r',a')  & \quad \text{if $r'<r$} \\
  \end{array} \right.  \hspace{0.5cm}
\eea

Now we consider the statistics calculated by binning cluster separations  in a bin of width $\Delta r$. In this case the pairwise velocity estimate is averaged over cluster pairs with separations within the finite bin,
\bea
\hat{\Vij}(r)\rightarrow \frac{1}{V_{bin}} \int_{r-\Delta r/2}^{r+\Delta r/2} \tilde{r}^2d\tilde{r}  \int d\Omega  \hat{\Vij}(\tilde{r})
\eea
where we again assume spherical symmetry.
Volume averaging over a bin of size $\Delta r=R_{\mathrm{max}}-R_{\mathrm{min}}$ yields
\bea
 j_1(kr) &\rightarrow& \frac{3}{R_{i,\mathrm{max}}^3-R_{i,\mathrm{min}}^3} \int_{R_{i,\mathrm{min}}}^{R_{i,\mathrm{max}}} r^2 \  j_1(kr) dr.
 \eea
 Using that
 \bea
  \int_{R_{i,\mathrm{min}}}^{R_{i,\mathrm{max}}} r^2 \  j_1(kr) dr &=& R_{i,\mathrm{min}}^3 \tilde{W}(k R_{i,\mathrm{min}}) - R_{i,\mathrm{max}}^3 \tilde{W}(k R_{i,\mathrm{max}})\\
  \eea
  with
  \bea
  \tilde{W}(x)=\frac{2 \cos(x) + x\sin(x)}{x^3}.
  \eea
  Binning in $r$ translates into replacing the Bessel function with a function related to the bin limits,
    \bea
   j_1(kr) &\rightarrow& \frac{3}{R_{i,\mathrm{max}}^3-R_{i,\mathrm{min}}^3} \left(R_{i,\mathrm{min}}^3 \tilde{W}(k R_{i,\mathrm{min}}) - R_{i,\mathrm{max}}^3 \tilde{W}(k R_{i,\mathrm{max}})\right)
 \equiv W_{\Delta}(kr).
 \eea
 Rewriting the volume averaged correlation function in terms of the power spectrum,
 \bea
 \frac{2 r}{r'^2} \bar{\xi}(r) &=&\frac{3}{r'^2 \pi^2} \int dk k P(k) j_1(kr),
 \eea
 and with
 \bea
 \frac{1}{r^2} \rightarrow \frac{ 4 \pi \Delta r}{V_{\Delta}(r)},
 \eea
 the full, angle-averaged covariance for the mean pairwise velocity, excluding measurement error, is given by the sum of a Gaussian cosmic variance and shot noise component   plus a Poisson  component,

\bea
&&C_{\Vij}(r,a,r',a')= C_{\Vij}^{\mathrm{Gaus.}}(r,a,r',a') + C_{\Vij}^{\mathrm{Poiss.}} (r,a,r',a')
 \eea
 with
 \bea
 C_{\Vij}^{\mathrm{Gaus.}}(r,a,r',a') &=& \frac{4}{\pi^2 V_s(a)} \frac{H(a)a}{1+\xi(r,a)}\frac{H(a')a'}{1+\xi(r',a')} f_g(a) f_g(a')\delta_{aa'} \int dk \left( P(k,a)+\frac{1}{n(a)} \right) \left( P(k,a')+\frac{1}{n(a')} \right) \nonumber \\
 &\times& W_{\Delta}(kr) W_{\Delta}(k r') \hspace{0.5cm} 
 \\ \nonumber \\
C_{\Vij}^{\mathrm{Poiss.}} (r,a,r',a') &=& \frac{4}{\pi^2 V_s(a) } \frac{H(a)a}{1+\xi(r,a)}\frac{H(a')a'}{1+\xi(r',a')}  f_g(a) f_g(a')\ \delta_{aa'}
  \left\{
\begin{array}{l l}
    \frac{\Delta r'}{n(a)^2 V_{\Delta}(r')} \int dk k P(k,a) W_{\Delta}(kr) & \quad \text{if $r'\geq r$ }\\
     \frac{\Delta r}{n(a')^2 V_{\Delta}(r)} \int dk k P(k,a') W_{\Delta}(kr') & \quad \text{if $r'<r$}. \\
  \end{array} \right. \nonumber 
  \\
\eea
These results are used in equation (\ref{eq:shot}) in the main text.

%%%%%%%%%%%%%%%%%%%%%%%%%%%%%%%%%%%%%%%%%%%%
% BIB
%%%%%%%%%%%%%%%%%%%%%%%%%%%%%%%%%%%%%%%%%%%%
\newpage
\bibliographystyle{apsrev}

\bibliography{big_bib.bib}

\end{document}